\pdfoutput=1

\PassOptionsToPackage{hyphens}{url}
\documentclass[camera,letterpaper,nomarginnotes,nonarrowgutter]{jpaper}

\usepackage{siunitx}
\usepackage{mwe}

\usepackage{cite}
\usepackage{amsmath,amsfonts}
\usepackage{amssymb}
\usepackage{graphicx}
\usepackage{textcomp}
\usepackage{dblfloatfix}
\usepackage{multirow}
\usepackage{booktabs}
\usepackage{balance}
\usepackage{array}
\usepackage{eucal}
\usepackage{wrapfig}
\usepackage{enumitem}
\usepackage{soul}
\newcommand{\Hl}[2][\empty]{\ifx#1\empty
\else
\sethlcolor{#1}\fi
\hl{#2}}
\usepackage{soul,color}
\soulregister\Hl{7}
\soulregister\ref7
\soulregister\cite7
\soulregister\pageref7

\usepackage{graphicx}
\usepackage{clipboard}
\usepackage{tikz}

\usepackage[many]{tcolorbox}

\usepackage[acronym,nonumberlist,nowarn]{glossaries}
    \glsdisablehyper
    \newacronym{ADI}{ADI}{Alternating Direction Implicit}
\newacronym{AGU}{AGU}{Address Generation Unit}
\newacronym[longplural={Access History Tables}]{AHT}{AHT}{Access History Table}
\newacronym{AHT-R}{AHT-R}{Access History Table - Read}
\newacronym{AHT-W}{AHT-W}{Access History Table - Write-Back}
\newacronym{AIP}{AIP}{Access Interval Predictor}
\newacronym{AL}{AL}{Added Latency for column accesses}
\newacronym{ASIC}{ASIC}{Application Specific Integrated Circuit}
\newacronym{AVX}{AVX}{Advanced Vector Extensions}
\newacronym[longplural={Basic Block Vectors}]{BBV}{BBV}{Basic Block Vector}
\newacronym{BHT}{BHT}{Branch History Table}
\newacronym{HBM}{HBM}{Hybrid Bandwidth Memory}
\newacronym{DDR4}{DDR4}{Double-Data Rate 4}

\newacronym{NN}{NN}{Neural Network}
\newacronym{BTB}{BTB}{Branch Target Buffer}
\newacronym{CAS}{CAS}{Column Address Strobe}
\newacronym{AMAT}{AMAT}{Average Memory Access Time}
\newacronym{CCD}{CCD}{Column to Column Delay}
\newacronym{AI}{AI}{Arithmetic Intensity}
\newacronym[longplural={Columnar Database Systems}]{CDBS}{CDBS}{Columnar Database System}
\newacronym[longplural={Chip Multiprocessors}]{CMP}{CMP}{Chip Multiprocessor}
\newacronym{CMT}{CMT}{Chip Multithreading}
\newacronym[longplural={Coarse-Grain Reconfigurable Arrays}]{CGRA}{CGRA}{Coarse-Grain Reconfigurable Array}
\newacronym{C-RAM}{C-RAM}{Computational-RAM}
\newacronym{CWD}{CWD}{Column Write Delay}
\newacronym{DBI}{DBI}{Dynamic Binary Instrumentation}
\newacronym[longplural={Database Management Systems}]{DBMS}{DBMS}{Database Management System}
\newacronym{DDR}{DDR}{Double Data Rate}
\newacronym{DEWP}{DEWP}{Dead Line and Early Write-Back Predictor}
\newacronym{DRAM}{DRAM}{Dynamic Random Access Memory}
\newacronym{DSBP}{DSBP}{Dead Sub-Block Predictor}
\newacronym{DSM}{DSM}{Decomposed Storage Manage}
\newacronym{FAW}{FAW}{Four row Activation Window}
\newacronym{FP}{FP}{Floating-Point}
\newacronym{EDP}{EDP}{Energy-Delay Product}
\newacronym{FCFS}{FCFS}{First-Come First-Serve}
\newacronym{FIFO}{FIFO}{Fist In, First Out}
\newacronym{FSM}{FSM}{Finite State Machine}
\newacronym{FPGA}{FPGA}{Field-Programmable Gate Array}
\newacronym[longplural={Functional Units}]{FU}{FU}{Functional Unit}
\newacronym{GAg}{GAg}{Global Adaptive branch prediction using one Global PHT}
\newacronym{GAs}{GAs}{Global Adaptive branch prediction using per-Set PHT}
\newacronym{GCC}{GCC}{GNU Compiler Collection}
\newacronym{GEMS}{GEMS}{General Execution-driven Multiprocessor Simulator}
\newacronym[longplural={General Purpose Processors}]{GPP}{GPP}{General Purpose Processor}
\newacronym{HMC}{HMC}{Hybrid Memory Cube}
\newacronym{HIVE}{HIVE}{HMC Instruction Vector Extensions}
\newacronym{IATAC}{IATAC}{Inter-Access Time per Access Count}
\newacronym{ILP}{ILP}{Instruction Level Parallelism}
\newacronym{IPC}{IPC}{Instructions per Cycle}
\newacronym{ISA}{ISA}{Instruction Set Architecture}
\newacronym{IRAM}{IRAM}{Intelligent RAM}
\newacronym{KIPS}{KIPS}{Kilo Instructions per Second}
\newacronym{LDS}{LDS}{Linked Data Structure}
\newacronym{LFSR}{LFSR}{Linear Feedback Shift Register}
\newacronym{LFMR}{LFMR}{Last-to-First Miss-Ratio}
\newacronym{MLP}{MLP}{Memory-Level Parallelism}

\newacronym{LLC}{LLC}{last-level cache}
\newacronym{LRU}{LRU}{Least Recently Used}
\newacronym{LSU}{LSU}{Load Store Unit}
\newacronym{LTP}{LTP}{Last-Touch Predictor}
\newacronym{LvP}{LvP}{Live-time Predictor}
\newacronym{LWP}{LWP}{Last Write Predictor}
\newacronym{MARSS}{MARSS}{Micro Architectural and System Simulator}
\newacronym{McPAT}{McPAT}{Multi-core Power, Area, and Timing}
\newacronym{MIPS}{MIPS}{Microprocessor without Interlocked Pipeline Stages}
\newacronym{MOB}{MOB}{Memory Order Buffer}
\newacronym{MMU}{MMU}{Memory Management Unit}
\newacronym{MPKI}{MPKI}{Misses per Kilo-Instruction}
\newacronym{MSHR}{MSHR}{Miss-Status Handling Registers}
\newacronym{NAS}{NAS}{Numerical Aerodynamic Simulation}
\newacronym{NMC}{NMC}{Near-Memory Computing}
\newacronym{NMP}{NMP}{Near-memory processing}
\newacronym{MCA}{MCA}{Memory-Centric Architecture}
\newacronym{NoC}{NoC}{network-on-chip}
\newacronym{NPB}{NPB}{NAS Parallel Benchmark}
\newacronym{NUCA}{NUCA}{Non-Uniform Cache Architecture}
\newacronym{NUMA}{NUMA}{Non-Uniform Memory Access}
\newacronym{OoO}{OoO}{Out-of-Order}
\newacronym{OpenMP}{OpenMP}{Open Multi-Processing}
\newacronym{OS}{OS}{Operating System}
\newacronym{PAg}{PAg}{Per-address Adaptive branch prediction using one Global PHT}
\newacronym{PAs}{PAs}{Per-address Adaptive branch prediction using per-Set PHT}
\newacronym{PC}{PC}{Program Counter}
\newacronym[longplural={Pointer-Chasing Engines}]{PCE}{PCE}{Pointer-Chasing Engine}
\newacronym{PCM}{PCM}{Performance Counter Monitor}
\newacronym{PIM}{PIM}{Processing-in-Memory}
\newacronym[longplural={Pattern History Tables}]{PHT}{PHT}{Pattern History Table}
\newacronym{RAPL}{RAPL}{Running Average Power Limit}
\newacronym{RAS}{RAS}{Row Address Strobe}
\newacronym{RAT}{RAT}{Registers Alias Table}
\newacronym{RC}{RC}{Row Cycle}
\newacronym{RCD}{RCD}{RAS to CAS Delay}
\newacronym[longplural={Row-based Database Systems}]{RDBS}{RDBS}{Row-based Database System}
\newacronym{ROB}{ROB}{Reorder Buffer}
\newacronym{RRD}{RRD}{Row to Row activation Delay}
\newacronym{RP}{RP}{Row Precharge}
\newacronym{RTL}{RTL}{Register Transfer Level}
\newacronym{RTP}{RTP}{Read To Precharge}
\newacronym{RVU}{RVU}{Reconfigurable Vector Unit}
\newacronym{SDP}{SDP}{Skewed Dead-Block Predictor}
\newacronym{HPC}{HPC}{high-performance computing}
\newacronym{MAC}{MAC}{multiply-accumulate}
\newacronym{SESC}{SESC}{Superescalar Simulator}
\newacronym{SSE}{SSE}{Streaming SIMD Extensions}
\newacronym{SFP}{SFP}{Spatial Footprint Predictor}
\newacronym{SiNUCA}{SiNUCA}{Simulator of Non-Uniform Cache Architectures}
\newacronym{SMT}{SMT}{Simultaneous Multi-Threading}
\newacronym{SIMD}{SIMD}{Single Instruction Multiple Data}
\newacronym{SPEC}{SPEC}{Standard Performance Evaluation Corporation}
\newacronym{SoC}{SoC}{System-on-Chip}
\newacronym{SPP}{SPP}{Spatial Pattern Predictor}
\newacronym{SRAM}{SRAM}{Static Random Access Memory}
\newacronym{SSOR}{SSOR}{Symmetric Successive Over-Relaxation}
\newacronym{SSV}{SSV}{Search Set Vector}
\newacronym{TLB}{TLB}{Translation Look-aside Buffer}
\newacronym{TLP}{TLP}{Thread Level Parallelism}
\newacronym[longplural={Through-Silicon Vias}]{TSV}{TSV}{Through-Silicon Via}
\newacronym{VWQ}{VWQ}{Virtual Write Queue}
\newacronym{WTR}{WTR}{Write Recovery time}
\newacronym{WR}{WR}{Write To Read delay time}
\newacronym[longplural={stencil processing units}]{SPU}{SPU}{stencil processing unit}

\usepackage{pifont}

\usepackage{wrapfig}

\usepackage{float}
\usepackage{listings}
\usepackage{graphicx}
\usepackage[newfloat, frozencache]{minted}
\usepackage[colorlinks,urlcolor=blue,linkcolor=blue,citecolor=blue]{hyperref}

\usepackage{cleveref}
\crefname{section}{§\hspace{-2pt}}{§§}
\Crefname{section}{§}{§§}

\newcommand{\mech}{{Casper}}  

\newcommand{\boldone}{\ding{202}}
\newcommand{\boldtwo}{\ding{203}}
\newcommand{\boldthree}{\ding{204}}
\newcommand{\boldfour}{\ding{205}}
\newcommand{\boldfive}{\ding{206}}

\newif\ifrevisionieee
\revisionieeefalse

\ifrevisionieee
\newcommand{\juangg}[1][0]{}
\else
\newcommand{\juangg}[1]{\textcolor{teal}{#1}}

\fi

\newif\ifrevision
\revisiontrue

\ifrevision
\newcommand{\jgl}[1][0]{}
\newcommand{\juang}[1][0]{}
\newcommand{\rbc}[1]{#1}
\else

\newcommand{\juang}[1]{\textcolor{teal}{#1}}
\newcommand{\rbc}[1]{{\color{blue}#1}}
\fi

\newcommand{\rbca}[1]{{\color{black}#1}}

\newif\ifsubmission
\submissiontrue

\ifsubmission
\newcommand{\juan}[1][0]{}

\usepackage{soul}
\newcommand{\gfeedback}[1][0]{}
\newcommand{\gagan}[1][0]{}
\definecolor{ddgreen}{rgb}{0.00, 0.50, 0.00}
\newcommand{\gs}[1]{\textcolor{black}{#1}}

\newcommand{\ad}[1][0]{}
\newcommand{\alain}[1][0]{}

\newcommand{\gf}[1][0]{}
\newcommand{\geraldo}[1][0]{}

\newcommand{\nasi}[1][0]{}
\newcommand{\nfeedback}[1][0]{}

\else
\newcommand{\jgl}[1]{\textcolor{magenta}{JGL: #1}}
\newcommand{\juan}[1]{\textcolor{magenta}{#1}}

\usepackage{soul}
\newcommand{\gfeedback}[1]{[\textcolor{red}{GAGAN: #1}]}
\newcommand{\gagan}[1]{\textcolor{red}{#1}}

\newcommand{\ad}[1]{\textcolor{green}{AD: #1}}
\newcommand{\alain}[1]{\textcolor{green}{#1}}

\newcommand{\gf}[1]{\textcolor{blue}{GF: #1}}
\newcommand{\geraldo}[1]{\textcolor{blue}{#1}}

\newcommand{\nasi}[1]{\textcolor{VioletRed}{#1}}
\newcommand{\nfeedback}[1]{[\textcolor{VioletRed}{Nastaran: #1}]}

\fi

\AtBeginDocument{\providecommand\BibTeX{{\normalfont B\kern-0.5em{\scshape i\kern-0.25em b}\kern-0.8em\TeX}}}

\def\BibTeX{{\rm B\kern-.05em{\sc i\kern-.025em b}\kern-.08em
    T\kern-.1667em\lower.7ex\hbox{E}\kern-.125emX}}

\newcommand{\boxbegin} {
	\begin{tcolorbox}[enhanced, frame hidden, colback=gray!50, breakable]
}

\newcommand{\boxend} {
	\end{tcolorbox}
}

\newcommand{\yboxbegin} {
	\begin{tcolorbox}[breakable, enhanced, frame hidden,
	enlarge top by=-0.25cm,
   enlarge bottom by=-0.1cm,
	colback=yellow!50]
}

\newcommand{\yboxend} {
	\end{tcolorbox}
}

\begin{document}

\title{Casper: Accelerating Stencil Computations\\ using Near-Cache Processing}


\author{Alain Denzler$^1$ \hspace{1em} Geraldo F. Oliveira$^1$ \hspace{1em} Nastaran Hajinazar$^{1,2}$ \hspace{1em} Rahul Bera$^{1}$ \\\vspace{0.2em} \hspace{-1em} Gagandeep Singh$^1$ \hspace{1em} Juan Gómez-Luna$^1$ \hspace{1em} Onur Mutlu$^1$\\
    \vspace{0.5em}
    $^1$ETH Zürich \hspace{1em} $^2$Simon Fraser University
}




\maketitle

\begin{abstract}

\rbc{Stencil {computations} {are commonly used} in a wide variety of scientific applications, ranging from large-scale weather prediction to solving partial differential equations.}
Stencil computations are \rbc{characterized by three properties}: (1) low arithmetic intensity, (2) limited temporal data reuse, and (3) regular and predictable data access pattern. As a result, stencil computations are typically bandwidth-bound workloads, which experience {only} limited benefits from the deep cache hierarchy of modern CPUs.

In this work, we propose \mech, a near-cache accelerator consisting of specialized stencil {computation} units connected to the last-level cache (LLC) of a traditional CPU. \mech~is based on two key ideas: (1) avoiding the cost of moving rarely reused data {throughout} the cache hierarchy, and (2) exploiting the regularity of the data accesses and the inherent parallelism of stencil {computations} to increase overall performance. 
With {small} changes in LLC address decoding logic and data placement, \mech~performs stencil computations at the peak \rbca{LLC bandwidth}. \rbc{We show that by tightly coupling lightweight stencil {computation} units near LLC, \mech~improves performance of stencil kernels by $1.65\times$ on average ({up to $4.16\times$}) compared to a commercial high-performance multi-core processor, while reducing {system} energy consumption by $35$\% on average ({up to $65\%$}). \mech~provides $37\times$ ({up to $190\times$}) improvement in performance-per-area compared to a state-of-the-art GPU.}
\end{abstract}



\section{Introduction}
\label{sec:introduction}

A stencil operation~\cite{gysi2015modesto} defines a computation pattern where elements in a multidimensional grid are updated based on the values of a fixed pattern of neighboring points. 
Computations using stencil operations (called \textit{stencil computations}) are a key building block of important~\gls{HPC} applications~\cite{colella2004defining} and are used in a wide range of workloads including climate modeling~\cite{fuhrer2018near}, seismic imaging \cite{mcmechan1983migration}, fluid dynamics~\cite{anderson1995computational}, and electromagnetic simulations~\cite{taflove1988review}. 
Stencil computations encompass a large percentage of the overall runtime of such applications~\cite{maruyama2014optimizing, fuhrer2013towards, olschanowsky2014study,opeanccdoc,mostafazadeh2018roofline}. For example, stencil computations represent over 90\% and 60\% of the overall runtime in {a} computational fluid dynamics solver~\cite{mostafazadeh2018roofline} and the COSMO climate simulation model~\cite{opeanccdoc, doms1999nonhydrostatic}, respectively. 
Consequently, a large body of research~\cite{christen2011patus, gysi2015modesto, datta2009optimization, strzodka2010cache, tang2011pochoir, ragan2013halide, meng2011performance, nguyen20103, henretty2011data, jaeger2012automatic, frigo2007memory, olschanowsky2014study, kamil2005impact, stengel2015quantifying, brandvik2010sblock, fuhrer2018near, armejach2018stencil, waidyasooriya2016opencl,de2018designing,sano2013multi,van2019coherently,singh2019narmada,chi2018soda,nero,li2019pims,waidyasooriya2019multi,cattaneo2015accelerate,yantir2020efficient,wester2014deriving} highlights the need for highly efficient stencil computations. 
However, the current compute-centric processing systems, such as multi-core CPUs and GPUs, fail to fully utilize their on-chip resources (e.g., deep cache hierarchy, high throughput floating-point engines) when computing stencil operations~\cite{datta2009optimization,nero}. This results in low performance and low energy efficiency for stencil computations in current systems. 
In this work, we show that a careful domain-specific hardware/software co-design can improve the performance and energy efficiency of stencil computations, at a much lower overhead than existing general-purpose solutions.

\begin{figure}[h]
    \centering
    \includegraphics[width=\linewidth]{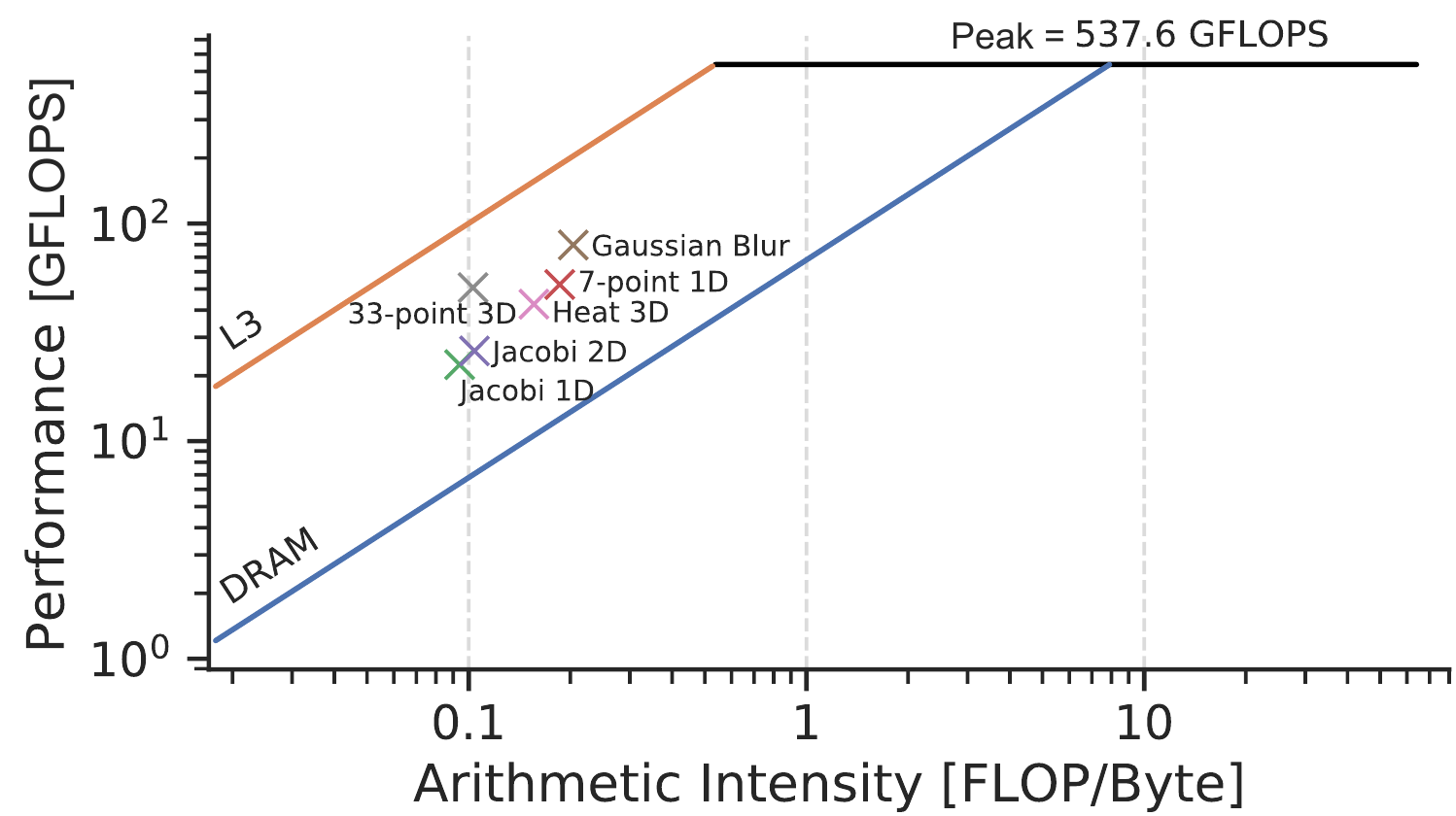}
    \caption{Roofline plot for a multi-core system~\cite{xeon-e7-4850, intelxeon} running six stencil kernels.}
    \label{fig:roofline}
\end{figure}

\textbf{Why a stencil accelerator?}
\autoref{fig:roofline} shows the roofline plot~\cite{williams2009roofline} of important stencil kernels on a server-class CPU {(a 16-core Intel Xeon CPU}~\cite{xeon-e7-4850, intelxeon}{)}. 
The horizontal line \gs{is} the peak floating-point performance of the system. 
{The stencil kernels are multithreaded and vectorized, and operate on double-precision floating-point values.} 
The DRAM and L3 lines show the peak \geraldo{main} memory and \gls{LLC} bandwidth, respectively. \geraldo{We make two observations. First,} the stencil kernels \emph{are not compute bound}, having low arithmetic intensity ranging from \SI{0.09}{ FLOP/B} to \SI{0.2}{ FLOP/B}. \geraldo{Second}, all the kernels are \emph{bounded by memory resources}, i.e., located on the left side of the inflection point and below the memory lines. More {precisely}, all the kernels are located below the L3 line and above the DRAM line, which shows that the stencils are bound by the \gls{LLC} bandwidth rather than the main memory bandwidth. 
Given these observations, we conclude that the high number of \gls{LLC} accesses is the main bottleneck for stencil computations, {which causes} such computations to experience only limited benefits from the deep cache hierarchies in modern CPU architectures.

Prior works \cite{augustin2009optimized, datta2008stencil, phillips2010implementing,singh2019low} also show that stencil kernels are bottlenecked by memory due to their low arithmetic intensity, leading to under-utilization of computational resources in compute-centric platforms (\rbca{e.g.,} CPU \rbca{and} GPU). 
These prior works demonstrate that stencil kernels can leverage {only} 21.8\% \cite{augustin2009optimized}, 
{46\%~\cite{szustak2019performance}, 34\%~\cite{szustak2019performance},}
and 46\%~\cite{datta2008stencil} of the computational resources of a multi-core CPU, {a 2-socket server-grade CPU, a 4-socket CPU, and a} GPU, respectively, even \nasi{in presence of} code optimizations (e.g., temporal blocking). 
{These percentages are inline with Figure}~\ref{fig:roofline}{, where all tested stencils achieve less than 20\% of the peak performance (537.6 GFLOPS).}
Therefore, existing general-purpose processors cannot deliver high performance and high {energy} efficiency for stencil computations, thus opening up the space for custom stencil-based accelerators~\cite{sano2013multi,van2019coherently,singh2019narmada,chi2018soda,nero,li2019pims,waidyasooriya2019multi,cattaneo2015accelerate,yantir2020efficient,wester2014deriving,armejach2018stencil,chi2018soda,christen2011patus,datta2009optimization,de2018designing,gysi2015modesto,henretty2011data,meng2011performance,strzodka2010cache,tang2011pochoir,waidyasooriya2016opencl,Singh2021trets}.

Processing-in-Memory (PIM) is a promising paradigm for
accelerating memory-bandwidth-bound workloads\gs{, which have} low arithmetic intensity~\cite{eckert2018neural, seshadri2017ambit,ahn2015pim,li2019pims,imani2019floatpim, mutlu2020modern,ghose.ibmjrd19,gomezluna2021arxiv, gomezluna2022IEEEAccess, oliveira2021damov, ghose2018enabling, mutlu2019processing}. The key idea of \geraldo{the} PIM paradigm is to move computation close to or even into the memory devices where the data resides (i.e., caches~\cite{Fujiki2019duality, eckert2018neural, aga2017compute,lockerman2020livia,subramaniyan2017cache,nag2019gencache,reduct,Pattnaik2019isca}, DRAM~\cite{ahn2015scalable,ahn2015pim, li2019pims, gomezluna2021arxiv, gomezluna2022IEEEAccess,kim2018grim,puli2018active,nai2017graphpim,gao2017tetris,seshadri2017ambit,hajinazarsimdram,oliveira2021damov,boroumand2019conda,lazypim,besta2021sisa,boroumand2018google,fernandez2020natsa,cali2020genasm, hashemi2016accelerating,singh2021fpga, nero, singh2019napel,seshadri2017simple,syncron,mutlu2019enabling,pattnaik2016scheduling,hsieh2016transparent,stone1970logic,gokhale1995processing,IRAM_Micro_1997,IBM_ActiveCube,oskin1998active,kang1999flexram,DIVA_2002,C_RAM_1999,imani2019floatpim,brockman2004low,farmahini2015nda,kim2016neurocube, zhang2014top, li2017drisa,deng2018dracc, angizi2019graphide,oliveira2017nim,drumond2017mondrian,santos2017operand, gu2020ipim,mutlu2019processing,ghose2018enabling,ghose.ibmjrd19,mutlu2020modern,seshadri2019dram, seshadri2017buddy, seshadri2015fast, boroumand2021google}, storage~\cite{yavits2021giraf,lee2017extrav,kang2021near,genstore,keeton1998case,acharya1998active,Shafiee2016,li2016pinatubo,Fujiki2018reram}), eliminating the need to move the data to the processor and resulting in {higher} performance and {lower} energy consumption. Stencil computations are a prime candidate for acceleration using {the} PIM \nasi{paradigm}. 
In this work, we explore the opportunity to improve the performance and energy efficiency of stencil computations in traditional multi-core CPUs {using computation} near the \gls{LLC}.

\textbf{Why near \gls{LLC}?} 
We exploit \gls{LLC} as the prime location for computation, as opposed to offloading {the computation} to {the} off-chip main memory~\cite{li2019pims,van2019coherently,singh2019narmada,yantir2020efficient,nero} or higher levels of caches (e.g., L2~\cite{lockerman2020livia}) for three {main} reasons. First, the per-thread {datasets} for stencil kernels in widely-deployed \rbca{scientific applications} are \rbca{typically} tiled to fit inside the \gls{LLC} of traditional workstation-class CPUs~\cite{cosmo_knl,fuhrer2018near}. Hence, placing computation {near} \gls{LLC} minimizes unnecessary data transfers {between} main memory and caches {and} cores. 
Second, the on-chip \gls{LLC} bandwidth is multi-fold {(e.g., about $10\times$ in Intel Xeon~\cite{intelxeon})} higher than {the} traditional DDR-based DRAM main memory bandwidth. Third, though computing in {higher}-level caches (e.g., L2) can theoretically provide higher bandwidth, {the additional \rbca{data transfers} {required by cache management protocols} (e.g., back invalidations, write backs, etc.) reduces the effective bandwidth significantly}. {Moreover, bringing data with low reuse (common in stencil computation data) to the higher-level caches results in major energy waste that can be alleviated by keeping such data in lower-level caches and performing the computation there.}

\textbf{Our goal} in this paper is to design a near-\gls{LLC} accelerator that improves the performance and energy efficiency of stencil computations by minimizing the unnecessary data movement \nasi{between the memory and CPU, and within the cache hierarchy}.

To this end, we propose \mech, a novel hardware/software codesign specifically targeted at stencil computations. We minimize data movement by placing a set of stencil processing  units (SPUs) near the \gls{LLC} of a traditional CPU architecture.
{\mech} provides novel mechanisms to {incorporate} data mapping changes and support unaligned loads needed for high-performance stencil computations. Computation is mapped to {SPUs} \rbca{in such a way} that each \gls{SPU} operates on the data that is located in the closest \gls{LLC} slice. This reduces the overall data access latency and energy consumption while matching the compute performance to the peak bandwidth of the \gls{LLC}.

Placing a stencil accelerator next to the \gls{LLC} of a CPU introduces two key challenges. The first challenge is to maximize {LLC} bandwidth utilization. To address this challenge, \mech~\geraldo{leverages} the notion of \emph{{streams}}~\cite{wang2019stream,cattaneo2015accelerate,khailany2001imagine,ciricescu2003reconfigurable,nowatzki2017stream} to expose the memory level parallelism that exists in the stencil computation to the \glspl{SPU}. Each stream represents a set of consecutive memory accesses with a fixed stride. The notion of {streams} enables \mech~to maximize memory bandwidth utilization without requiring complex structures {(e.g., those that ii{exist} in high-performance cores to perform dynamic instruction ii{scheduling}~\cite{mutlu2003runahead,mutlu2003runahead2,mutlu2005techniques})}.

The second challenge is to minimize the data movement between different cache slices. In stencil computations, the neighboring grid points need to be accessed to compute the stencil operation for each grid point. However, current systems employ an address mapping scheme that distributes data over different \gls{LLC} slices and provides load balance and fairness across CPU cores~\cite{zohouri2018combined}. 
Such a mapping scheme can potentially map neighboring grid points to different \gls{LLC} slices, introducing data transfers over the \gls{NoC} and thereby eliminating the benefits of near-cache computing. To address this challenge, \mech~{uses a new hash function to align the data mapping to the grid structure of the stencil computation and place neighboring grid points in the same slice of the \gls{LLC}}.

\rbc{We evaluate \mech~using six widely used stencil kernels with up-to 3-dimensional grid domains. \mech~outperforms a commercial multi-core CPU, on average, by $1.65\times$ ({up to $4.16\times$}) and reduces the energy consumption by $35\%$ ({up to $65\%$}). Compared to a state-of-the-art GPU, \mech~improves performance-per-area\gs{, on average,} by $37\times$ ({up to $190\times$}).}

\vspace{0.5em}
\noindent We make the following \gs{key} contributions:
\begin{itemize}
    \item We propose \mech, {the first} near-cache accelerator for stencil computations. 
\mech~addresses the memory bottleneck in stencil computations by {(1)~eliminating the need to move data to the ii{processing core} for computation, (2)~{minimizing} the data movement within the cache hierarchy, and (3)~maximizing the utilization of the \gls{LLC} bandwidth. \mech~achieves this with an area overhead of less than 1\% ii{for 16 SPUs in a Marvell ThunderX 2~\cite{thunderx2}, a server-class ARM CPU}}.

\item {We present a \gagan{memory}-centric execution model that maximizes the LLC bandwidth utilization by exposing the memory level parallelism that exists in the stencil computation to the near-cache accelerators.}
\item {We provide hardware support to minimize data movement between different cache slices using a new mapping scheme that improves spatial locality for stencil data.}
    \item We evaluate the effectiveness of \mech~{using six widely used stencil kernels and demonstrate that \mech~outperforms a commercial multi-core CPU, on average, by $1.65\times$ ({up to $4.16\times$}) and reduces the energy consumption by $35\%$ ({up to $65\%$}). Compared to a state-of-the-art GPU, \mech~improves performance-per-area\gs{, on average,} by $37\times$ ({up to $190\times$}).}
    
\end{itemize}

\vspace{-2mm}
\section{Background}
 \subsection{Stencil Computations}\label{sec:jacobi-2d-bg}

Stencil computations \cite{gysi2015modesto} update {the} points in a data grid based on a fixed pattern of neighboring points. This fixed pattern, called the \textit{stencil}, is applied on the complete grid iteratively until either a convergence criterion or a fixed number of steps are reached. Stencil computations are widespread in scientific computing, and are considered one of the \rbca{thirteen} dwarfs of scientific computing \cite{colella2004defining}.

\begin{figure}[th]
    \centering
    \includegraphics[width=\columnwidth]{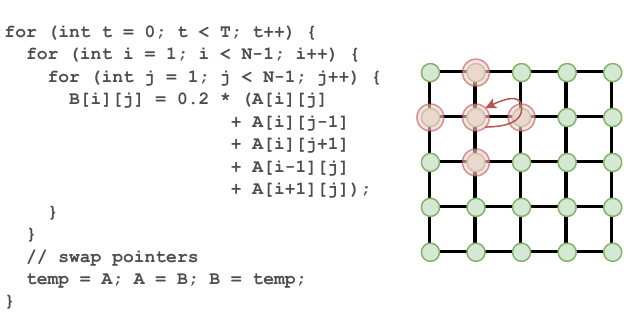}
    \vspace{-8mm}
    \caption{2D Jacobi stencil pseudo-code and data access pattern.}
    \label{fig:jac2d}
\end{figure}

Stencil computations exhibit several common properties. We explain these properties using \gs{a} \textit{Jacobi} stencil~\cite{demmel1997applied} {that is} commonly used to solve discretized partial differential equations, as an example. \autoref{fig:jac2d} shows the source code and data access pattern of a 2-dimensional Jacobi stencil. The computation performs arithmetic mean of each point in the grid and its immediate neighbors in all directions. This implementation \geraldo{uses} three nested loops where the outermost loop iterates over time steps and the two inner loops sweep over the complete 2D grid. From this example, we can identify four common properties of stencil computations. First, the computation is embarrassingly parallel because the read and write data sets are disjoint. Second, the computation is regular and statically analyzable. The data dependencies and the structure of the computation can be analyzed ahead-of-time and do not depend on {a} dynamic input. 
Third, the arithmetic intensity of the stencil is low. Fourth, only a few types of operations are needed to compute the stencil. For example, in case of \textit{Jacobi} stencil, only a floating-point \gls{MAC} operation is performed for each \juang{input} grid point (e.g., multiply \texttt{A[][]} by 0.2 and accumulate) when computing an output grid point (\texttt{B[][]}). 
These properties {make stencil computations a suitable candidate for high-performance acceleration (due to the highly-parallel and regular computation) while making them a natural fit for near-memory acceleration (due to the low arithmetic intensity and relying on only few types of operations).}

\subsection{Memory-Centric Architectures: Overview and Limitations}
\label{subsec:background-mca}

Processing-in-Memory (PIM) architectures (e.g.,~\cite{Fujiki2019duality, eckert2018neural, aga2017compute,lockerman2020livia,subramaniyan2017cache,nag2019gencache, reduct,ahn2015scalable,ahn2015pim, li2019pims, gomezluna2021arxiv, gomezluna2022IEEEAccess,kim2018grim,puli2018active,nai2017graphpim,gao2017tetris,seshadri2017ambit,hajinazarsimdram,oliveira2021damov,boroumand2019conda,lazypim,besta2021sisa,boroumand2018google,fernandez2020natsa,cali2020genasm, hashemi2016accelerating,singh2021fpga, nero, singh2019napel,seshadri2017simple,syncron,mutlu2019enabling,pattnaik2016scheduling,hsieh2016transparent,stone1970logic,gokhale1995processing,IRAM_Micro_1997,IBM_ActiveCube,oskin1998active,kang1999flexram,DIVA_2002,C_RAM_1999,imani2019floatpim,brockman2004low,farmahini2015nda,kim2016neurocube, zhang2014top, li2017drisa,deng2018dracc, angizi2019graphide,oliveira2017nim,santos2017operand, gu2020ipim,mutlu2019processing,ghose2018enabling,ghose.ibmjrd19,mutlu2020modern,yavits2021giraf,lee2017extrav,kang2021near,genstore,keeton1998case,acharya1998active,Shafiee2016,li2016pinatubo,Fujiki2018reram,Olgun2021PiDRAM,Kim2019drange,Olgun2021isca,Mutlu2021date}) attempt to address the memory bottleneck issue by performing computation in proximity to the memory and thereby reducing the overheads of {data movement in the} system. There are two approaches to PIM~\cite{mutlu2020modern}: (1) processing-using-memory (PuM), which performs computation inside the memory array itself, using the analog {operational} properties of the memory cells~\cite{hajinazarsimdram, li2019pims, gomezluna2021arxiv, gomezluna2022IEEEAccess,puli2018active,seshadri2017ambit,hajinazarsimdram,oliveira2021damov,besta2021sisa,singh2021fpga,nero,seshadri2017simple,syncron,mutlu2019enabling,stone1970logic,gokhale1995processing,IRAM_Micro_1997,IBM_ActiveCube,oskin1998active,kang1999flexram,DIVA_2002,C_RAM_1999,imani2019floatpim,brockman2004low, zhang2014top, li2017drisa,deng2018dracc,seshadri2013rowclone,angizi2019graphide,oliveira2017nim,santos2017operand, gu2020ipim,mutlu2019processing,ghose2018enabling,ghose.ibmjrd19,mutlu2020modern,seshadri2019dram, seshadri2017buddy,seshadri2015fast,Fujiki2019duality, eckert2018neural, aga2017compute,lockerman2020livia,subramaniyan2017cache,nag2019gencache,reduct, yavits2021giraf,lee2017extrav,kang2021near,genstore,keeton1998case,acharya1998active,Shafiee2016,li2016pinatubo,Fujiki2018reram,chang2016low,seshadri2019dram,seshadri2016processing,Olgun2021PiDRAM,Kim2019drange,Olgun2021isca}; and (2) processing-near-memory (PnM), which employs compute elements close to the memory arrays, for example on the logic layer of a 3D-stacked DRAM device~\cite{boroumand2019conda,boroumand2018google,fernandez2020natsa,singh2019napel,hsieh2016transparent,kim2016neurocube,hashemi2016micro,Qiuling2013,Pugsley2014,farmahini2015nda,pattnaik2016scheduling,Akin2015isca,Hwan2015pact,Mingyu2016hpca,Ping2016isca,seshadri2015gather,Zhiyu2017,Gao2017pact,Morad2015GPsimd,nai2017graphpim,Gwangsun2015sc,Xi_2015,Mingu2014ICASSP,drumond2017mondrian,cali2020genasm,hashemi2016accelerating,ahn2015scalable,ahn2015pim,lazypim,asghari2016chameleon,kim2018grim,gao2017tetris,Dai2018,Huang20202,Youwei2019}.

\juan{The PuM proposals, either in DRAM~\cite{seshadri2017ambit,hajinazarsimdram,besta2021sisa,seshadri2017simple,li2017drisa,deng2018dracc,angizi2019graphide,seshadri2013rowclone,seshadri2015fast,chang2016low,seshadri2017buddy,seshadri2019dram,seshadri2016processing,Kim2019drange} or SRAM~\cite{Fujiki2019duality, eckert2018neural,aga2017compute,Mingu2014ICASSP}, perform bulk bitwise and arithmetic (e.g., addition, multiplication, reduction) operations.} 
Even though these works benefit from the large internal bandwidth provided by the memory device (DRAM or SRAM), they require the application to follow a rigid data layout and data mapping scheme to align the operands correctly within the memory arrays. Such data layout and alignment management \juan{are} not straightforward and \juan{remain} an open research problem. In contrast, \juan{our work} introduces hardware and software optimizations to efficiently orchestrate the data movement or handle unaligned operands.

PnM proposals span a wide range of applications{, such as} 
neural networks~\cite{reduct, gao2017tetris, oliveira2017nim,boroumand2021google}, databases~\cite{drumond2017mondrian, santos2017operand,Xi_2015}, mobile workloads~\cite{boroumand2018google}, bioinformatics~\cite{kim2018grim,Alser2020ieeemicro}, image processing~\cite{gu2020ipim}, and graph processing~\cite{ahn2015scalable,lockerman2020livia,nai2017graphpim,Youwei2019}. 
\juan{PIMS}~\cite{li2019pims} proposes \juan{a PnM approach (in the logic layer of \gls{HMC}~\cite{hmc2013hybrid})} to accelerate stencil operations.
The proposed solution leverages the computational capabilities that are already present in \gs{an} \gls{HMC} device\footnote{The \gls{HMC} specification~\cite{hmc2013hybrid} defines a series of 8 to 16-byte atomic operations that can be executed directly within the memory device.} to execute {the} addition {operation} that is commonly used in stencil computations. Despite the performance improvement compared to a CPU-centric model, {PIMS} \rbca{suffers} from {two important} shortcomings. First, prior works \cite{rosenfeld2014performancepaul, oliveira2017generic} show that the atomic operations provided by the \gls{HMC} device can only exploit a small fraction of the total internal memory bandwidth, {creating} a bottleneck for the bandwidth-hungry stencil {computations}. Second, {due to the limited area and power budget available inside the logic layer of \gls{HMC},} PIMS {only} supports the addition operation {and continues to rely on the host processor to execute other more complicated operations such as multiplication}. {Therefore}, PIMS still incurs a high memory traffic between the host and the memory device to compute a single stencil operation.

To our knowledge, {\mech} is the first work to tightly integrate specialized compute units into the \gls{LLC} of a \rbca{traditional} CPU to {accelerate} stencil computations. In contrast to prior works, \mech~accounts for the \emph{fundamental} properties of stencil computations, such as data access pattern, aiming to provide a hardware/software co-design that can fully exploit the high memory bandwidth provided by the cache, while efficiently orchestrating the data movement required to execute the stencil operation.

\vspace{-2mm}
\section{\mech: Overall Architecture}
\label{sec:HLD}
\glsreset{SPU}
\glsreset{LLC}

We propose \mech, a novel near-cache accelerator, to improve the performance and energy efficiency of stencil computations. This section introduces the high-level overview of our architecture (Section~\ref{subsec:HLD-arch}), the execution model of \mech~(Section~\ref{subsec:HLD-exec}), and stencil processing unit (Section~\ref{subsec:HLD-SPU}).

\subsection{Overview}
\label{subsec:HLD-arch}

\mech~is a near-cache accelerator that leverages {a  memory-centric  execution  model and hardware support} to accelerate stencil operations. A typical shared \gls{LLC} is partitioned into multiple cache slices, connected through {the network on chip (\gls{NoC})}. \mech's hardware is composed of a set of \glspl{SPU}, placed in each cache slice of the \gls{LLC}. In addition to accessing the local cache slice, each \gls{SPU} can load data from remote \gls{LLC} slices through the \gls{NoC}. Similar to a regular \gls{LLC}, \gls{SPU}s can load data from other levels of the cache hierarchy into the \gls{LLC}. 

Computing at the \gls{LLC}'s peak throughput is only possible {when each} \gls{SPU} in the system {loads} data from \gs{{its} local}  \gls{LLC} slice, thus avoiding overheads related to \rbca{NoC} traffic. However, the existing mapping of memory addresses to the \gls{LLC} slices does not guarantee mapping consecutive cache lines to the same slice. In fact, while mapping scheme information remains undisclosed, prior work shows that consecutive cache lines are usually mapped to different \gls{LLC} slices in order to provide fairness and load balancing across CPU cores \cite{yarom2015mapping}. Due to the streaming nature of stencil computations and the dependency on a small {group of neighboring grid points}, scattering data through the cache slices would increase the number of remote data requests {made by each} \gls{SPU}, penalizing performance and energy consumption. {To address this issue}, \mech~maps blocks of consecutive memory addresses to the same \gls{LLC} slice. {To this end}, we introduce hardware support for a \textit{stencil segment} {i.e., a contiguous region of physical memory that contains the data set of the stencil {kernel}}. \mech~customizes the mapping of memory addresses to \gls{LLC} slices for {the stencil} segment {in order to} optimize data locality when computing stencils. Data mapped outside the stencil segment follows the {conventional} address mapping {scheme}.

\subsection{Execution Model}
\label{subsec:HLD-exec}

At a high\gs{-}level, each \gls{SPU} performs the stencil operation on each grid point sequentially. The points are accessed in row-major order, following their layout in memory. For each grid point, the \gls{SPU} loads the input data from \gls{LLC}. Once the data arrives at the \gls{SPU}, the \gls{SPU}'s execution unit performs a single multiply operation with a predefined constant, whose output is added to the accumulator. The final result is stored in the results array after all computations for the grid point have been performed. Then, the \gls{SPU} proceeds to calculate the next grid point. 

To accelerate this process, \mech~abstracts data movement through a series of \emph{stream} operations, which {enable} feeding data to the executing units at a high {throughput}. A \emph{stream} is a sequence of data elements of the same type located in consecutive physical memory addresses. \nasi{Each stream is} characterized by a start address, data type width, a number of elements, and a position pointer. When iterating over a stream, initially, a position pointer \nasi{indicates} the start of \gs{a} stream. Upon receiving a control signal, this pointer is incremented to point to the stream's next element. The stream can then be advanced until the final element in the stream is reached. \nasi{The stream} abstraction facilitates iterating over \nasi{stencil} data {by simply incrementing the position pointer, as opposed to} loading each data element using absolute memory addresses. Since a stencil computation moves through {the entire} grid at the same pace, maintaining the same relative offsets for the stencil pattern, it is \nasi{naturally} possible to use streams to abstract {such an} access pattern. {Accordingly}, all the streams can be advanced at the same pace, marshaled by the \gls{SPU} moving through the grid. 

\subsection{Stencil Processing Unit}
\label{subsec:HLD-SPU}

\begin{figure}[h]
    \centering
\includegraphics[width=\columnwidth]{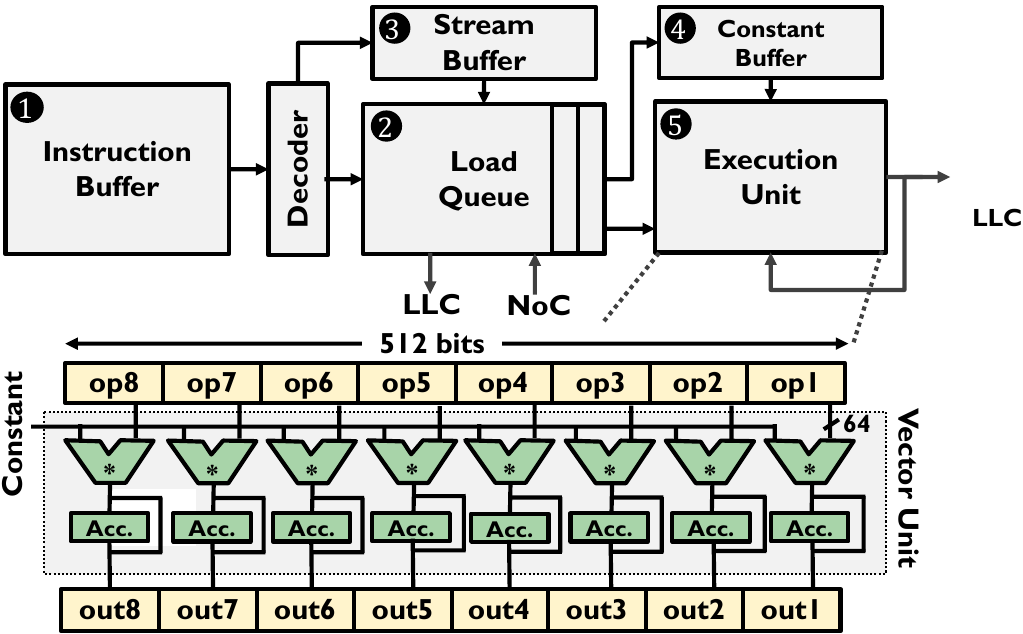}
    \vspace{-2mm}
    \caption{Components of one \gls{SPU}.}
    \label{fig:spu}
\end{figure}

\autoref{fig:spu} shows the architecture of a \gls{SPU}. The main building blocks {of an \gls{SPU}} are {\boldone~the instruction buffer, \boldtwo~the load queue, \boldthree~the stream buffer, \boldfour~the constant buffer, and \boldfive~the execution unit}. The complete \gls{SPU} is pipelined to maintain single-cycle instruction throughput. The \gls{SPU} controls the complete computation, from instruction fetch until retire. Therefore, it does not \nasi{require} any interaction with the CPU.

\textbf{Instruction Buffer.} The instruction buffer has a capacity to hold $64$ \emph{compressed} instructions to compute a stencil. Every instruction encodes the operands and the control signals {necessary} for one type of stencil operation. The same sequence of instructions is applied to every stencil grid point due to the regular nature of the stencil computation. We introduce the instruction set \nasi{used in \mech} in Section~\ref{sec:interfaces}.

\textbf{Load Queue.} The load queue is responsible for issuing data requests to the memory subsystem. While the \gls{SPU} processes instructions in-order, \rbc{the varying memory access latency} \geraldo{of the memory subsystem} can result in responses that arrive out-of-order. The load queue acts as a buffer to hold the data until all the previous longer-latency memory requests have been completed. This buffer space ensures that the {correct} instruction order is maintained. The load queue is sized to hide the latency of accessing the {LLC}'s local slice because as it is the preferred location for fetching the data.

\textbf{Stream and Constant Buffers.} The stream buffer holds the current state of the streams. For every decoded instruction, this state is loaded from the stream buffer, and the effective address is calculated to issue the memory requests to the \gls{LLC}. The constant buffer holds the constant factors needed for the \gls{MAC} operation. The stream and constant buffers are initialized by API calls \geraldo{(Section~\ref{sec:api})} before the \gls{SPU} starts the stencil computation.

\textbf{Execution Unit.} 
As shown in \autoref{fig:spu}, the execution unit of a \gls{SPU} is comprised of \rbc{a $512$-bit vector unit which operates on vectors of $8$ double-precision floating-point elements}. 

For every input grid point, the \gls{SPU} loads the data elements from the neighboring points, multiplies them with a constant factor, and accumulates the results. \rbc{This final accumulated result is then stored in the output grid point}. Therefore, each \gls{SPU} execution unit only computes one kind of instruction: a double-precision floating-point \gls{MAC} operation.

\section{\mech: Micro-architectural Support}\label{sec:hardware_support}

\label{lab:R1/2}\Copy{R1/2}{{The design and implementation of Casper require us to solve several key system integration challenges, as we discuss in this section. 
First, the SPUs access frequently data from two cache lines. We introduce two simple changes to the LLC row decoder for efficient access (Section}~\ref{sec:unaligned_loads}{). 
Second, the cache lines accessed by an SPU should preferably reside in the same LLC slice. We introduce lightweight remapping to ensure that contiguous blocks of stencil data map to the same LLC slice (Section}~\ref{sec:llc_data_mapping}{). 
Third, \mech~should be compatible with existing cache coherency mechanisms (Section~}\ref{sec:coherence_support}{). 
Fourth, \mech~should be able to concurrently run with other CPU processes (Section}~\ref{sec:context_switch}{).}}

\subsection{Supporting Unaligned Loads}\label{sec:unaligned_loads}
\begin{figure}[ht]
    \centering
    \includegraphics{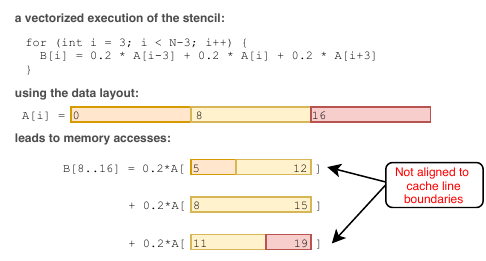}
    \caption{Unaligned loads occurring during the vectorized execution of a stencil.}
    \label{fig:shiftexample}
\end{figure}

While the stream abstraction offers efficient memory accesses {for \glspl{SPU}}, the \gls{LLC} architecture only supports data accesses {that are} aligned to {the} \SI{64}{B} {cache line} boundaries. As the relative offsets used in streams are not necessarily aligned to cache line boundary, the loaded data might need to be realigned before it can be used by \gls{MAC} compute.
For example, in \autoref{fig:shiftexample} each \gls{SPU} computes {eight} $64$-bit output grid points (\texttt{B[i]}) for a $3$-point Jacobi-1D stencil.
\rbc{The access {to} the center point of the stencil \texttt{A[i]} is correctly aligned to the \SI{64}{B} boundary {of the cache line} (shaded in yellow) such that the data for this point can be used for computation as soon as it arrives at the \gls{SPU}. However, to gather several input grid points at indices $+3$ (\texttt{A[i+3]}) and $-3$ (\texttt{A[i-3]}), the data coming from the cache needs to be correctly shifted (shaded in orange and red), and two cache lines need to be combined to assemble the operands for the computation. As a result, preparing the operand for the \gls{MAC} unit involves two loads, one shift, and one combine operation on the data from the two cache lines. 
These additional operations {to resolve} unaligned data accesses lead to} {two main inefficiencies: (1) underutilization of MAC units, and (2) cache bandwidth overhead. 
The reason is that the number of load/store operations per MAC operation increases. For example, the sequential code shown at the top of} \autoref{fig:shiftexample} {needs $4$ load/store operations per $3$ MAC. However, the vectorized execution of this stencil (bottom part of} \autoref{fig:shiftexample}{) would need $6$ load/store operations per $3$ MAC, i.e., two cache line loads for \texttt{A[5]} to \texttt{A[12]}, one cache line load for \texttt{A[8]} to \texttt{A[15]}, two cache line loads for \texttt{A[11]} to \texttt{A[19]}, and one cache line store for the elements of \texttt{B}.}


To address the above challenges, we introduce two simple modifications to the \gls{LLC} row decoding logic {to} (1)~support loading data aligned to any \SI{8}{B} boundary, and (2)~correctly align data for \gls{SPU} execution. 
{These two modifications allow us to load values from two adjacent cache lines (e.g., values \texttt{A[5]} to \texttt{A[12]} in} \autoref{fig:shiftexample}{) in one access.} 
Our modified \gls{LLC} row decoding logic reduces \gls{SPU}'s complexity and area footprint {by avoiding the need to add} a large register file and/or extra logic for shifting and packing the partial cache lines inside the \gls{SPU}.

\textbf{Implementation Challenges.}
Loading data aligned to \SI{8}{B} boundaries {can} potentially {result in} loading data located on two cache lines. This introduces two key challenges. \rbc{First, \emph{two} tags need to be matched to locate the two cache lines {in one access (i.e., using one address)}. Second, the correct data from {each of the two} cache lines must be loaded using only one {provided} address.}

\rbc{To address the first challenge, we enable the cache to match two tags in parallel by adding a second read port to the tag array.} Since the two cache lines involved in an unaligned load are always at consecutive addresses {(e.g., values \texttt{A[5]} to \texttt{A[12]} in} \autoref{fig:shiftexample}{)}, they are guaranteed to be mapped to different cache sets, and thus, there are no conflicts during the tag matching process. If at least one cache line is not in a readable state, a regular cache miss is generated, and the request is stalled until the {miss} resolves.

\rbc{To address the second challenge, we make changes into the \gls{LLC} row decoding logic to load data either from the requested address or one of the cache line{s} located inside the 64B-vicinity of the requested address. More specifically, we add one 3-to-1 multiplexer for each \gls{LLC} SRAM row. The inputs to the multiplexer are set to the row decoder output \nasi{of} the current row and both the adjacent rows. The multiplexer selects the appropriate row(s) based on the row selection signal.}

{We explain next the solutions to both challenges.}

\begin{figure}[h]
\centering
    \includegraphics[width=\columnwidth]{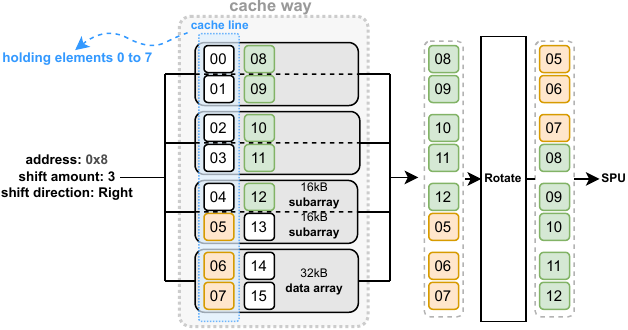}
\caption{Loading unaligned data from the {LLC}. The two cache lines involved in the access are mapped to the same cache way. One cache way consists of 4$\times$\SI{32}{kB} data arrays, each having 2$\times$\SI{16}{kB} subarrays.}
    \label{fig:unaligned_load}
\end{figure}

\textbf{Loading Shifted Cache Lines.}
\autoref{fig:unaligned_load} shows the execution sequence of {an} unaligned access to the \gls{LLC}. In this example, we consider the request to {the} cache line holding elements 8 to 15, shifted to the right by three positions. This corresponds to the access to input grid points \texttt{A[i-3]} in \autoref{fig:shiftexample} {(i.e., values \texttt{A[5]} to \texttt{A[12])}}. 
\juang{{First, we consider} the case where the} two cache lines involved in the access are mapped to the \juang{\emph{same}} cache way. One cache way consists of 4$\times$\SI{32}{kB} data arrays, each having 2$\times$\SI{16}{kB} subarrays. One \SI{16}{kB} subarray holds 64 consecutive bits of a cache line for all the 2048 sets. Thus, each 64-bit segment of a cache line is stored in a different SRAM subarray. As shown in \autoref{fig:unaligned_load}, all the elements required to build the {requested} \SI{64}{B} block of data are stored in different subarrays of a single cache way \juang{(orange and green elements on the left part of the figure)}. Therefore, by selecting the correct data at the subarray level \juang{(using shift direction $shdr$ and amount $shamt$)}
it is possible to load all the elements using only one command, while maintaining the same throughput as a regularly aligned load {(any extra latency is negligible, since there are no conflicts during the matching of the two tags, and pipelined accesses hide it)}. 
To \juang{complete the unaligned access}, we need to rotate the data such that the correct element is at the first position of the \SI{64}{B} block of response data. To accomplish this, we use a rotate network that performs this operation before the final output. {This approach provides the \glspl{SPU} with the ability to load data that is aligned to arbitrary 8B boundaries} from the \gls{LLC}.

If the two cache lines involved in an unaligned load are mapped to \juang{\emph{different}} ways, the load sequence is similar. As the data and tag access happen in parallel inside the cache, the data load is initiated on all cache ways before the way hit has been confirmed. Thus for any access, all the ways present the data for the requested set on their output bus. Based on the way hit, the data from the correct way is selected for the output. We modify the way hit selection to select each subarray's output independently, depending on the shift amount and direction {that is provided} with the request.

\begin{figure}[!ht]
    \centering
    \includegraphics[width=0.8\columnwidth]{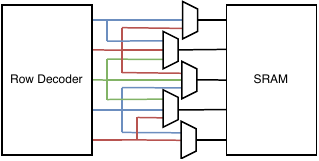}
\caption{Additional logic added between the row decoder and the SRAM array. All the multiplexers receive the same select signal.}
    \label{fig:decoder}
    \vspace{-1mm}
\end{figure}

\textbf{Row Decoding.} Locally at each SRAM array, the shift amount and {the} direction are used to determine whether to load the data from the requested address or {one of the two rows immediately adjacent to it}. {To this end, we include a local selection signal that reroutes the row selection signal to the correct row. \autoref{fig:decoder} shows the layout of the added logic for local selection signal which consists of one 3:1 multiplexer for each SRAM row. The inputs} to each multiplexer are the row decoder output for the current row and the signals for both of the adjacent rows. The multiplexer then selects which input to forward. If the subarray needs to select elements from the cache lines specified in the request's address, {the multiplexer} forwards the middle signal (the row activation does not change). If the element from the adjacent cache line should be loaded, the output depends on the shift direction. If the shift direction is to the right (left), the row at index -1 (+1) should be activated. This shifts the row selection signal by one position in the requested direction, and loads data stored on the adjacent cache line. We include logic at the edge of each subarray to compute the select signal for the multiplexers based on the shift direction, shift amount, and the subarray ID (i.e., the position of the bits stored in this subarray within a complete cache line). All the multiplexers for one SRAM subarray are controlled using the same select signal.

\subsection{LLC Data Mapping}\label{sec:llc_data_mapping}
Sliced \glspl{LLC} employ address mapping schemes that aim to provide load balance and fairness across CPU cores. {These mapping schemes use} hash-functions that are not disclosed by CPU vendors\cite{yarom2015mapping}, but are shown to map consecutive cache lines to different \gls{LLC} slices. This is a challenge for \mech, since the performance of the \gls{SPU} can only be maximized if the \gls{SPU} mainly accesses data stored on the {local} \gls{LLC} slice. {The computation} for one data point depends on input data from its neighboring grid points {which as consecutive cache lines can be mapped to different \gls{LLC} slices. This leads to significant} data transfers over the \gls{NoC}, thus reducing the benefits of near-cache computing.

To address this challenge and enable maximum performance for each \gls{SPU}, we introduce a mechanism to adapt the {mapping of data to \gls{LLC} slices to the needs of stencil computation}. The {new data mapping scheme} maps blocks of consecutive memory addresses to the same slice, allowing neighboring grid points to be stored in the same slice of the \gls{LLC}. Thus, remote slice accesses are reduced to only accesses to points that lie on the other side of the boundaries between the blocks.

We enable remapping {of} the data used for the stencil computation without affecting the mapping for other system data by introducing a \textit{stencil segment}. Following the proposal by~\cite{basu2013efficient}, a physically contiguous block of memory is used by the system to hold stencil data. {\mech~ensures that the physically contiguous blocks of data in the stencil segment are mapped to the same \gls{LLC} slice.} 
At each \gls{NoC} injection point, the system checks {whether ii{or not} the address is part of the stencil segment. In case the address belongs to a stencil segment a new hash function is ii{applied} to issue a stencil segment memory request to the \gls{LLC}}. Otherwise, the {conventional} hash function is applied. By deciding which mapping function to use based on a memory request's physical address, each address is mapped to exactly one cache slice, regardless of whether or not it contains stencil data.

{Sizing blocks to map stencil data to ii{the} LLC comes with a trade-off.} Smaller blocks introduce more remote slice accesses for multidimensional grids, which hurts performance. The boundary elements are all located on separate cache lines, and cannot be loaded using the unaligned load mechanism {as} the affected cache lines are stored on separate \gls{LLC} slices. On the other hand, smaller blocks allow partitioning physically contiguous arrays across these blocks and thus distributing the data evenly to cache slices. Therefore, data for the same grid points from different source arrays can map to the same slice. This mapping \nasi{scheme} improves locality for multi-source stencils at the expense of more remote slice accesses on the blocks' boundary. For example, in a 16 core system, data aligned to a 2MB boundary is mapped to the same slice, and the system can consume data from up to 16 arrays for one grid point from the local slice. {In this work, we design the hash function to map stencil data as a linear hash that statically maps contiguous blocks of 128kB to \gls{LLC} slices in round-robin fashion.\footnote{{128kB blocks provide a good tradeoff across our evaluated stencils. We leave the design of a configurable hash function for future work.}}} 

\subsection{Coherence Support}\label{sec:coherence_support}
An {SPU} loads data from the {LLC} by directly injecting a load/store request in the {LLC} controller's request queues. \mech~does not impact the current cache coherency mechanism, {as} requests from the {SPU} are injected into the same request queues as conventional requests from the CPU cores and the private caches. If a write request from an SPU targets a cache line that is not in writable state in the LLC, the coherency mechanism will trigger necessary state transitions and invalidations to allow the write to complete.

\subsection{Concurrent Execution with CPU and Context Switching}\label{sec:context_switch}
To enable concurrent execution with the CPU, we reserve one way of the {LLC} for other applications running in the CPU, similar to prior work \cite{eckert2018neural}. Additionally, high priority is assigned to the {SPU} process to minimize the occurrence of context switches. If a context switch happens, the state of the {SPU} remains unchanged as it is not allowed to start a new stencil computation before the current computation finishes.

\section{\mech: Programming Model Support}\label{sec:interfaces}

\subsection{\mech~ISA} 
\begin{figure}[ht]
    \centering
\includegraphics[width=\columnwidth]{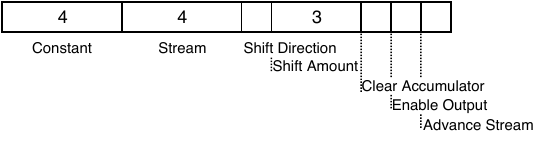}
\caption{Instruction layout for stencil computation.}
    \label{fig:microcode}
\end{figure}

\mech~makes use of specialized instructions to compute each point involved in the stencil operation. 
{\autoref{fig:microcode} shows the layout of a \mech~instruction.} \rbc{Every \mech~instruction is $15$-bits wide and comprises of (1) $4$b constant buffer index, (2) $4$b stream buffer index, (3) $1$b shift direction, (4) $3$b shift amount, and (5) $3$b control bits.} 15-bits instructions lead to a compact stencil code in \mech.\footnote{Common complex stencils have input sizes in the order of 30-40 points.} 
Note that \mech~reuses the instructions for all the grid points, thus reducing the instruction count.

After decoding a new instruction, the \gls{SPU} uses the 4-bits present in the constant field to \nasi{index the constant buffer (a small SRAM buffer)}, loading the requested double-precision constant value that is used by the execution unit during the computation.  The \gls{SPU} uses the 4-bits in the stream field to index the stream buffer and find the memory address of the requested stream. The shift direction and the shift amount fields are used to assemble the correct memory address that point to the appropriate data. {The final 3 bits are control bits} \emph{clear accumulator}, \emph{enable output}, and \emph{advance stream} {that} are used to reset the accumulators in the execution unit, enable the contents of the accumulator to be stored into the results array, and to advance the stream pointers, respectively.

We also provide a programming library that allows the user to easily generate \mech~instructions from user-level code. The generated instructions to execute a given stencil computation are then stored in a contiguous array that holds all the stencil instructions. This array is then stored in the instruction buffer of all the {SPU}s. In this paper, we statically analyze stencil operations and generate the appropriate set of \mech~instructions using our library. However, this step could be fully automated by a compiler due to stencil operations' regular nature.

\subsection{\mech~API}
\label{sec:api}
\label{lab:R1/3}\Copy{R1/3}
{\autoref{table:api} shows \mech~API functions. Calls to API functions are mapped directly to ISA instructions, which are broadcasted to all SPUs. 
These instructions are integrated into the existing ISA {(e.g., x86}~\cite{saini1993design}{, ARM}~\cite{jaggar1997arm}{, RISC-V}~\cite{waterman2016design}{)} using spare instruction opcodes. {We use the ARM ISA in our implementation (Section}~\ref{subsec:methodology-setup}{).}
Similar to other offload-based accelerators like GPUs, the CPU is not allowed to modify stencil data while the SPUs are running, to avoid corrupting the data. Enforcing this policy is the responsibility of the programmer.}

\begin{table*}[h!]
    \centering
    \caption{\mech~Programmer API}
    \scriptsize
    \begin{tabular}{m{10em}m{11.5em}m{45em}}
    \toprule
    \textbf{Function} & \textbf{Parameters} & \textbf{Description} \\
    \midrule
    \texttt{\textbf{initStencilSegment}} & \texttt{int \textit{size}} & Requests a physically contiguous memory region of the specified size from the system to hold the stencil data \\
    \midrule
    \texttt{\textbf{initStencilcode}} & \texttt{addr \textit{A}, int \textit{length}} & Takes a pointer to the microcode and the length of the code. After generating the code with helper functions provided by the programming library, the code is then broadcast to the \gls{SPU} \\
    \midrule
    \texttt{\textbf{initConstant}} & \texttt{double \textit{const}, int \textit{index}} & Iinitializes constant values that will be used during the multiplication step of a stencil operation. The function sets the specified constant value at the given index in the constant buffer \\
    \midrule
    \texttt{\textbf{initStream}} & \texttt{addr \textit{A}, int \textit{streamID}, int \textit{accID}} & Configures the streams used in the stencil code. The streams are configured per {SPU} to enable the programmer to tune the data layout to the grid's structure and minimize the number of off-slice cache accesses \\
    \midrule
    \texttt{\textbf{setNElements}} & \texttt{int \textit{n}, int \textit{accID}} & To {communicate} to each SPU how many elements to compute. All the SPUs maintain a counter to keep track of their progress and stop when the desired number of elements has been computed \\
    \midrule
    \texttt{\textbf{startAccelerator}} & - & The final function starts with the execution of all the {SPU}. After calling the start function, one of the \glspl{SPU} acts as the \juang{leader} and maintains a state to track the progress of all \glspl{SPU}. Once all the \glspl{SPU} finish their computation, the \juang{leader} signals the completion to the CPU \\
    \bottomrule
    \end{tabular}
    \label{table:api}
\end{table*}

\section{An Example: Jacobi-2D Stencil}

\begin{figure}[ht]
    \centering
    \includegraphics[width=\columnwidth]{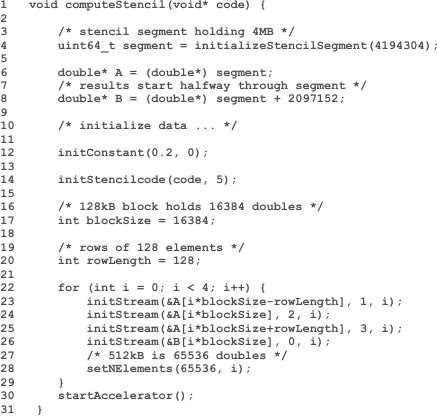}
    \vspace{-0.5em}
    \caption{Program code for Jacobi 2D stencil.}
    \label{fig:code}
\end{figure}

We illustrate how to program \mech~using the code in \autoref{fig:code} as an example. The code implements the Jacobi-2D stencil presented in Section \ref{sec:jacobi-2d-bg}. We explain the example using a system consisting of 4 \gls{LLC} slices and \gls{SPU}s.

First, a stencil segment covering \SI{4}{MB} is allocated (line 4). Then, we define the start of the arrays \texttt{A} and \texttt{B} such that the same grid point of both arrays is mapped to the same {LLC} slice (lines 6-8). 
The programmer then initializes the arrays with the stencil data. Furthermore, the constant for the multiplication (line 12) and the stencil instructions (line 14) are sent to the \gls{SPU}.
The streams for all the \gls{SPU} are configured inside the loop (lines 22-29). In this example, four streams are configured: three input streams to load the elements at \texttt{A[j-1][i]}, \texttt{A[j][i]} and \texttt{A[j+1][i]} (lines 23-25), and one output stream to store the result (line 26). As the elements \texttt{A[j][i-1]}, \texttt{A[j][i]}, and \texttt{A[j][i+1]} are laid out in consecutive memory addresses, we can reuse the same stream and leverage the support for unaligned loads by shifting the access to the left (right) by one element to load the data. Finally, the number of elements to compute for each \gls{SPU} (line 28) is configured, and the computation starts.

\begin{figure}[ht]
    \centering
    \includegraphics[width=\columnwidth]{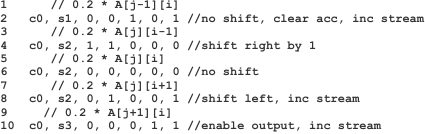}
    \vspace{-4mm}
    \caption{Instruction sequence for the Jacobi-2D stencil.}

    \label{fig:mc}
\end{figure}

\autoref{fig:mc} shows the stencil instructions executed on the {SPU}. As the stencil loads data from five input points, the instruction sequence consists of five instructions. All the input points are multiplied by the constant $0.2$. This means all the instructions encode the same constant factor c0. The instructions load data from three different streams: the first stream points to the value \texttt{A[j-1][i]}, aligned to a cache line boundary. The values \texttt{A[j][i-1]}, \texttt{A[j][i]}, and \texttt{A[j][i+1]} are stored in consecutive memory addresses. These three accesses all use the same stream, but using a shift amount and direction included in the request for the elements at indices \texttt{i+1} and \texttt{i-1}. This loads the correctly aligned data using the unaligned load mechanism. Finally, the third stream is configured to load the value at \texttt{A[j+1][i]}.

{As mentioned earlier (Section~\ref{sec:interfaces}),} the final 3 bits of each \mech~instruction hold control information for the {SPU}. {The first bit, i.e., clear accumulator bit, must be set by the first instruction of a grid point (line 2). The next bit is the enable output bit which is set in the last instruction of the sequence (line 10) and generates a store request. The final bit or the advance stream bit signals the {SPU} to advance the stream pointer and has to be set in the last instruction consuming data from each stream (lines 2, 8, 10)}.

\section{Methodology}

\subsection{Experimental Setup}\label{subsec:methodology-setup}

We simulate the performance of \mech~using ii{the} gem5 simulator {(v20.0)}~\cite{binkert2011gem5, lowe2020gem5} in syscall emulation mode. \autoref{tab:config} describes our system configuration in details. We use the ARM ISA and the Ruby memory model. The system is based on a generic modern server-class CPU, consisting of 16 out-of-order cores and three levels of cache, having a sliced {LLC} with \SI{2}{MB} per slice. Our baseline configuration uses the same system configuration without the \gls{SPU} and the \gls{LLC} changes we propose. Also, we include stride prefetchers at all levels of the cache hierarchy. We evaluate the performance and energy benefit of \mech~against the baseline CPU architecture, an NVIDIA Titan V GPU\cite{titanV}. We use an energy model based on CACTI 7.0 \cite{balasubramonian2017cacti} and energy models proposed by prior works~\cite{tsai2018adaptive, tsai2017jenga}.
We use the area model presented in \cite{shao2014aladdin} scaled down to 22nm to estimate the area of the {SPU}.\footnote{{We conservatively scale down the model as analyzed in}~\cite{salehi2015energy}{.}} For the GPU performance/area comparisons, we use the complete die size of 815mm² of the Titan V \mbox{\cite{nvidiaTitanDie}} because typical GPU-accelerated systems also need a host CPU to perform the computation. As a result, the total area for the end-user is either the complete GPU or the \mech~hardware modifications, added to the area of the existing host CPU.

\begin{table}[ht]
\caption{Simulation Parameters { for the baseline CPU and \mech}}
\resizebox{\columnwidth}{!}{\begin{tabular}{l l}
    \toprule
    Component & Configuration \\
    \midrule
    \textit{\mech} & 16 \glspl{SPU}, 1 SIMD unit/{SPU} (512-bits wide) \\
         & 10 entry load queue, \SI{0.016}{\nano\joule}/instruction \\
    {\textit{CPU}} & 16 out-of-order cores, 2 GHz, 8-wide issue, \\
         & 72 entry load queue, 64 entry store queue, \\
         & 1 SIMD unit/core (512-bits wide) \\
         & 224 entry ROB, \SI{0.08}{\nano\joule}/instruction \\
    \textit{L1I/D Cache} & \SI{32}{kB}, private, 8-way, 16 MSHRs, \\ 
         & 4 cycle round-trip latency, 2 load ports, 1 store port \\
         & 15/33 pJ per hit/miss \cite{tsai2018adaptive} \\
    \textit{L2 Cache} & \SI{256}{kB}, private, 8-way, 16 MSHRs, \\
         & 12 cycle round-trip latency, 1 load port, 1 store port \\
         & 46/93 pJ per hit/miss \cite{tsai2018adaptive} \\
    \textit{L3 Cache} & \SI{32}{MB}, shared, 16-way, 16 slices, 32 MSHRs/slice \\
         & 36 cycle round-trip latency, 1 load/store port per slice \\
         & 945/1904 pJ per hit/miss \cite{tsai2018adaptive} \\
    \textit{Coherence Protocol} & MESI \\
    \textit{Replacement Policy} & LRU replacement \\
    \textit{Hardware Prefetchers} & Stride prefetchers at all levels of the cache \\
    \textit{On-Chip Network} & mesh, XY-routing, \SI{64}{B}/cycle per direction \\
    \textit{Main Memory} & \SI{16}{GB}, DDR4, 4 channels, 160nJ per read/write~\cite{tsai2017jenga} \\
    \bottomrule
\end{tabular}}
\label{tab:config}
\end{table}

\subsection{Benchmarks}

We evaluate \mech~on six stencil benchmarks, including up to 3-dimensional stencils with varying data reuse characteristics, ranging from 3 input points (Jacobi 1D) up to 33 points for the 33-point 3D stencil. \alain{We use Jacobi 1D, -2D, 7-point 3D (heat diffusion) from Polybench~\cite{pouchet2012polybench}, a 5$\times$5 Gaussian blur filter \cite{canny1986computational}, the 7-point 1D kernel from~\cite{holewinski2012high}, and a 33-point 3D kernel to represent higher-order scientific simulations\cite{datta2008stencil,datta2009auto}.} All the benchmarks use Jacobi-style stencils with disjoint read and write data sets. 
{They all operate on double-precision floating-point values.} 
These benchmarks are elementary stencils that can be conjugated to form more complex stencils occurring in real-world applications. In addition to the standard data set sizes (which fit inside the {LLC} of the CPU), we also evaluate \mech~on data sets that (1) exceed the size of the {LLC} (named \textit{DRAM}), and (2) fit within the private L2 caches of the CPU (\textit{L2}). \autoref{tab:gridsizes} lists the domain sizes. 
In the appendix, \autoref{tab:app_instr} {shows the dynamic instruction counts of all evaluated stencils and datasets for the baseline CPU and for \mech.}

\begin{table}[h]
    \centering
    \footnotesize
     \caption{Domain size used for evaluations}
\begin{tabular}{m{3em}m{5em}m{6em}m{6em}}
        \toprule
        \textbf{Level} & \textbf{1D} & \textbf{2D} & \textbf{3D} \\
        \midrule
\textit{L2} & 131,072 & $512\times256$ & $64\times64\times32$\\
        \textit{L3} & 1,048,576 & $1024\times1024$ & $128\times128\times64$\\
        \textit{DRAM} & 4,194,304 & $2048\times2048$ & $256\times256\times64$\\
        \bottomrule
    \end{tabular}
\label{tab:gridsizes}
\end{table}

\section{Results}\label{sec:results}

{This section analyzes the performance, the energy consumption, and the hardware cost of \mech, and compares to the baseline CPU, the baseline GPU, and a PnM accelerator for stencil operations. 
The appendix includes} \autoref{tab:app_cycles} {and} \autoref{tab:app_energy}{, which contain our detailed measurements.}

\subsection{Performance}\label{sec:results-perf}

\begin{figure}[ht]
    \centering
    \includegraphics[width=\linewidth]{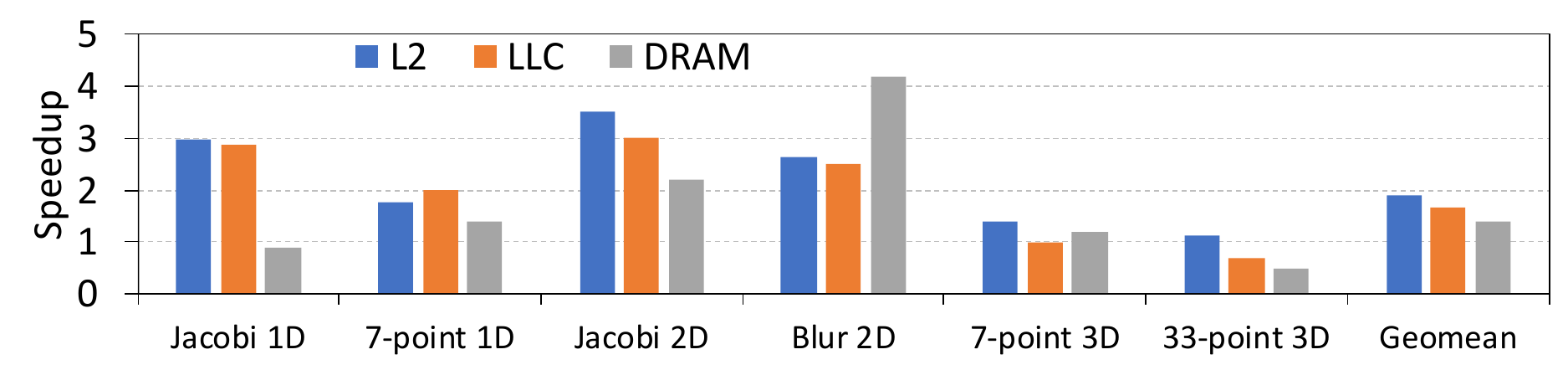}
    \caption{Speedup compared to the baseline multi-core system.}
    \label{fig:performance}
\end{figure}

\autoref{fig:performance} shows the speedup of \mech~compared to a 16-core CPU baseline system. For the datasets that fit within the {LLC}, which represent typical data set sizes for stencil computations, we observe an average speedup of 1.65$\times$. The 1- and 2-dimensional stencils achieve speedups between 1.66$\times$ and 3.0$\times$. However, the 3-dimensional stencils cannot achieve the same performance and even experience a slowdown in the 33-point stencil case. The reason for this performance {loss} is twofold. First, 3-dimensional stencils need to load a significant part of their input data from remote {LLC} slices, which introduces longer access latencies and lowers the throughput of the {SPU}. Second, the 33-point 3D stencil has good L1 cache behavior in the baseline, achieving a hit-rate of 95\%{, making it a good fit for} execution on a traditional CPU. We conclude that the performance benefits of \mech~are larger on lower-dimensional stencils that load most of their input data from the local {LLC} slice.

The average performance improvement for the {smaller data sets that fit} {in} the L2 caches of the CPU {(i.e., L2 in \autoref{fig:performance})} is 1.89$\times$. Even though the data is stored closer to the core and does not need to travel through the entire cache hierarchy, the speedups are similar to the larger data sets that fit {in} the {LLC}. This is {due to the fact that} the access latency from CPU to the L2 cache is similar to the latency between \mbox{{SPU}} and the closest \mbox{{LLC}} slice (12 vs 8 cycles load-to-use). 

\label{lab:R1/6}\Copy{R1/6}{Our results show that for large data sets that exceed the size of the {LLC} {(i.e., DRAM in \autoref{fig:performance})}, \mech~improves performance by 1.4$\times$, on average. The highest speedups are achieved by the 2-dimensional stencils, with Blur 2D reaching 4.16$\times$. We explain this by the fact that the baseline {CPU implementation of Blur 2D} has a very low \mbox{{LLC}} hit-rate (only 2\%), and thus the number of main memory accesses is 4$\times$ higher {compared to} \mech. The main reason for this low hit-rate {of the baseline CPU implementation} is that the prefetchers are interfering with demand accesses, evicting cache lines out of the \mbox{{LLC}} before they are used. The remaining stencils {(i.e., Jacobi 1D, 33-point 3D)} perform similar to the baseline or even experience slowdowns. The reason for this is the fact that the main memory bandwidth is the main bottleneck. The 33-point 3D stencil experiences a slowdown because it is well-suited for computation with smaller private caches for each core. We conclude that even though \mech~cannot alleviate the main memory bandwidth bottleneck {in the case of} large data sets, it does not lead to significant slowdowns for most stencils.}

\subsection{Energy Consumption}

\begin{figure}[ht]
    \centering
    \includegraphics[width=\linewidth]{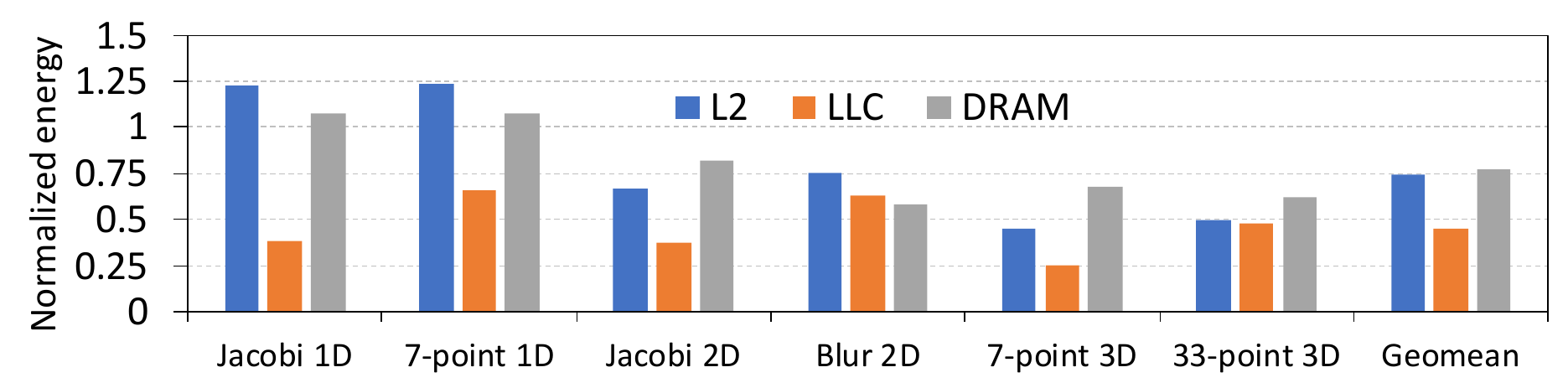}
    \caption{Normalized energy consumption compared to the 16-core baseline.}
    \label{fig:energy}
\end{figure}

\autoref{fig:energy} shows the normalized energy consumption of \mech~compared to the baseline CPU. For the data sets that fit into the {LLC} {(i.e., LLC in \autoref{fig:energy})}, \mech~reduces energy consumption by 55\%, on average. \mech~reduces energy consumption by 49\% even for the 33-point 3D stencil, whose performance is slower in \mech~than in the baseline. The reduction in energy consumption is {larger} for simpler stencils (Jacobi 1D/2D, 7-point 3D), reaching up to as 65\% for 7-point 3D. {This is due to the fact} that complex stencils perform more {LLC} accesses. {Furthermore,} since the \glspl{SPU} are situated close to the {LLC} slice, accessing {LLC} is not as energy-efficient as accessing the smaller L1 cache. This results in lower energy savings for the more complex stencils because of higher L1 reuse in baseline CPUs.

For {both} the smaller and larger data sets {(i.e., L2 and DRAM in \autoref{fig:energy})}, \mech~reduces energy consumption by 26\% and 23\%, on average respectively. We make two observations. First, \mech~increases the energy consumption of  the 1-dimensional benchmarks (Jacobi 1D and 7-point 1D) when compared to the baseline, for both small and large data sets. For the large data sets, this is the case because the CPU cores can be idle for most of the runtime, waiting for memory. For the smaller data set, the baseline's energy consumption is very low because there are very few {LLC} accesses. Since \mech~loads the data from the shared {LLC}, it increases energy consumption in such cases. Second, for all the other benchmarks, \mech~reduces energy consumption significantly. \mech~is more energy-efficient even for the benchmarks that the CPU baseline outperforms since our {SPU} design is more energy-efficient than the CPU baseline. Thus, we conclude that \mech~achieves significant reductions in energy consumption compared to a traditional out-of-order CPU.

\subsection{Comparison with GPU}
\begin{figure}[h]
    \centering
    \includegraphics[width=\linewidth]{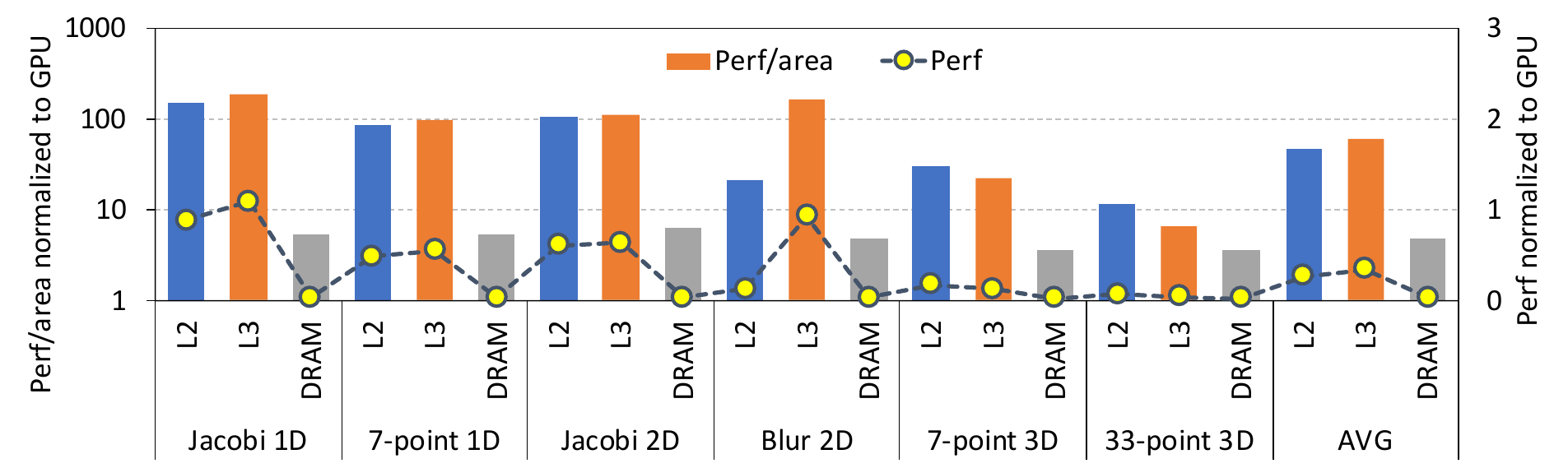}
    \caption{Performance/area compared to an NVIDIA Titan V GPU for three different domain sizes.}
    \label{fig:performance-gpu}
\end{figure}

\autoref{fig:performance-gpu} shows \mech's performance and performance-per-area normalized to an NVIDIA Titan V GPU. 
\rbc{GPU outperforms \mech~by $3.71\times$, $2.89\times$, and $36.64\times$ on average across all stencils that fit inside L2, LLC, and DRAM respectively.
However, in all stencil kernels, \mech~provides {higher} performance-per-area ({up to} 190$\times$ {compared to GPU}).}
{The reason for this large performance-per-area improvement is that 16 SPUs occupy 349$\times$ less area than the Titan V, i.e., $16 \times 0.146 mm^2$ (see Section}~\ref{sec:results-area}{) versus $815 mm^2$).} 
We observe that the L2- and LLC-sized data sets achieve performance-per-area improvements of 47$\times$ and 60$\times$, respectively. For these data set sizes, \mech~has the advantage of its tight integration into the large LLC. At the same time, the data does not fit into the GPU caches. For the large DRAM-sized data sets, however, the average improvement in performance/area is only 4.78$\times$. In this case, GPU improves relative performance due to higher main memory bandwidth.

\subsection{Comparison with PIMS}

PIMS~\cite{li2019pims} proposes a PnM accelerator targeting stencil operations. PIMS accelerator leverages the atomic operations available in the \gls{HMC} architecture to compute addition instructions in a stencil. Since PIMS represents the closest related work to \mech, we compare both accelerators using our stencil kernels. {To evaluate the performance of} PIMS, we {conservatively} consider {only} the latency of executing the atomic add operations, without accounting for the extra overhead {of (1)~loading} the results back from the \gls{HMC} device{, and (2)~executing} the multiply operations required by each stencil on the host CPU. We use as a reference in our {analysis} the peak throughput of the \gls{HMC} atomic operations reported by prior work~\cite{oliveira2017generic}.

\begin{figure}[h]
    \centering
    \includegraphics[width=\columnwidth]{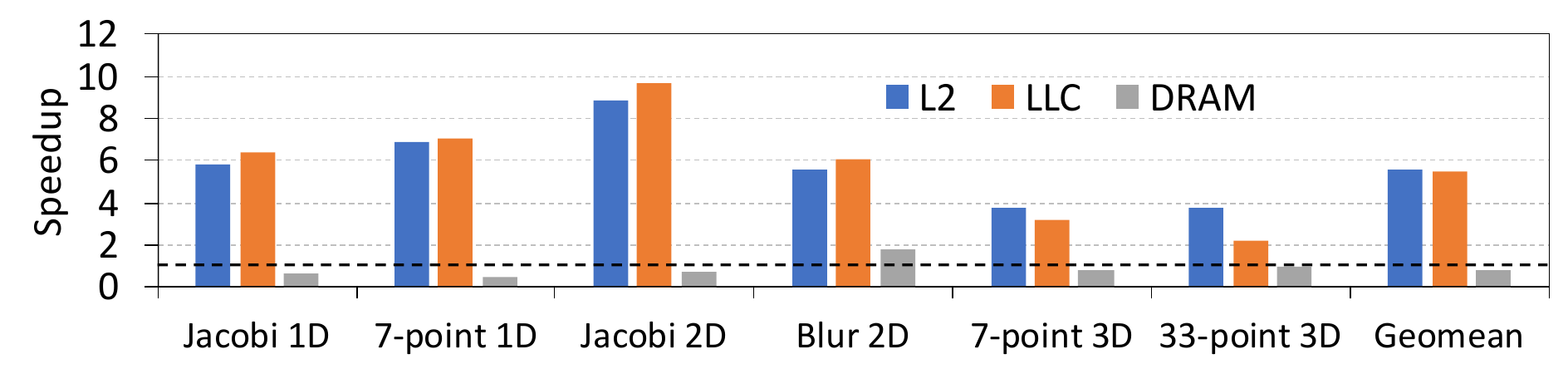}
\caption{Speedup compared to PIMS~\cite{li2019pims}.}
    \label{fig:performance-pims}
\end{figure}

\autoref{fig:performance-pims} shows the speedup of \mech~in comparison to PIMS. We make the following observations. First, \mech~provides an average speedup of 5.5$\times$ (up to 10$\times$) compared against PIMS, for the data set sizes that fit inside the on-chip caches. This happens because of the low throughput of the atomic operations HMC provides, which becomes the main bottleneck. Second, the stencils that do not fit into the caches perform worse using \mech~compared to PIMS. We attribute this to the fact that computing on the logic layer of the HMC has much higher memory bandwidth than the off-chip memory bus connected to the CPU for these bandwidth-bound stencils. We conclude that \mech~performs significantly better than PIMS on typical stencil datasets, \juang{i.e., those that fit within the LLC}.

\subsection{Effect of Individual Optimizations}

\begin{figure}[h]
 \centering
   \includegraphics[width=\linewidth]{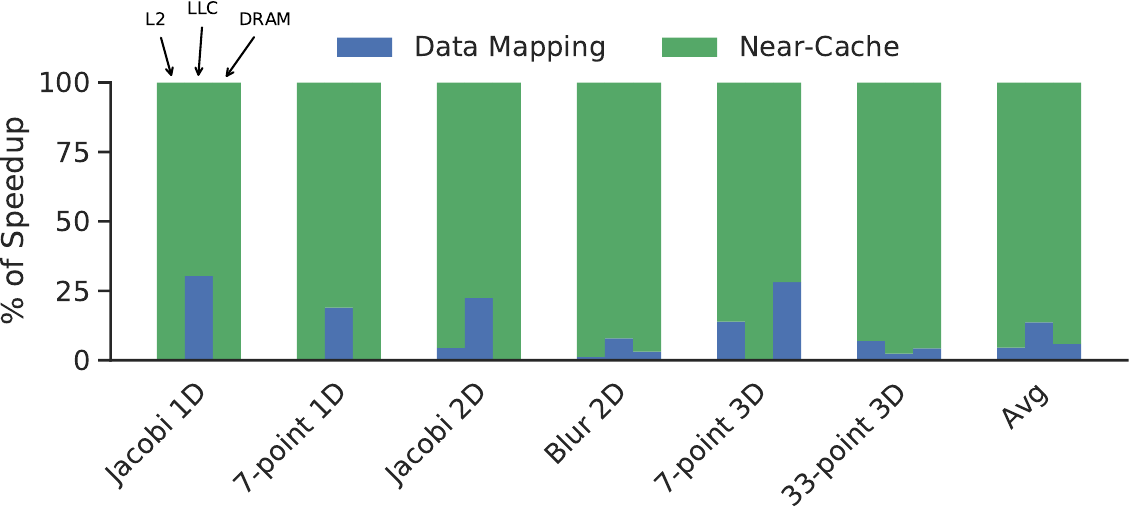}
  \caption{Contribution of custom data mapping and near-cache SPU location to the speedup of \mech~over a baseline system with SPUs near L1 for L2-/LLC-/DRAM-sized data sets.}
  \label{fig:performance-breakdown}
\end{figure}

\juang{The SPUs in \mech~take advantage of two key optimizations: (1)~a custom \emph{data mapping} in LLC, {ii{which} improves performance by increasing the locality of stencil data in the LLC and reducing the need for SPUs to access remote LLC slices}, and (2)~the \emph{near-cache} (near-LLC) location of the SPUs, which minimizes data access latency and leverages the peak bandwidth of the LLC. 
In this section, we evaluate the contribution of each {of these two} optimizations to the overall performance {of \mech}. 
The baseline for this analysis is a system where the SPUs are located next to the private L1 caches of CPU cores. The baseline LLC data mapping {conventionally} places consecutive cache lines in consecutive LLC slices (similar to~\cite{yarom2015mapping}). 
First, we apply {only} the data mapping optimization and compare the {\mech~against the} baseline. 
{Next}, we apply both the data mapping optimization and the near-cache optimization.}
{Then, we normalize to 100\% the speedup that results from the two optimizations together, and obtain the percentage that comes from the data mapping and from the near-cache location.} \autoref{fig:performance-breakdown} {shows bars with blue and green parts. The blue part represents the percentage of the speedup due to the data mapping, while the green part represents the percentage of the speedup due to the near-cache location.}

\juang{We make two observations from the results in \autoref{fig:performance-breakdown}. 
First, computing near-cache (green portion of the bars) is the major contributor to the speedup. 
Second, the custom data mapping (blue portion of the bars) produces up to 30\% of the speedup (Jacobi 1D with LLC data set), but its effect is negligible (or even negative) in several cases (1D and 2D benchmarks with L2/DRAM data sets, 7-point 3D with LLC data set). In these cases, the custom data mapping results in a number of accesses to remote LLC slices which is similar to the baseline data mapping.}

\subsection{Hardware Cost}\label{sec:results-area}

\noindent\textbf{Stencil Processing Unit.}
The area of one {SPU} scaled to \SI{22}{\nano\meter} technology is 0.146 mm\textsuperscript{2}. The most significant contributors to this area are the execution unit and the SRAM array used to buffer complete memory requests. 

\noindent\textbf{Unaligned Loads.}
Our hardware mechanism to support unaligned loads consumes an additional 0.14 mm\textsuperscript{2} compared to the baseline \SI{2}{MB} SRAM cache slice. This amounts to a 5\% increase in area per {LLC} slice. Almost the complete area overhead is attributed to the second read port of the tag array, which consumes 0.12 mm\textsuperscript{2} of space. The remaining hardware overhead of one 3:1 multiplexer per SRAM row, the barrel shifter to rotate the final output, and the split multiplexers for way selection are minimal compared to the tag array overhead.

\noindent\textbf{Address to {LLC} Slice Mapping.}
Identifying the stencil segment {requires} two registers to store the start and the length of the segment. The {address} comparison needs one adder and one comparator. The new mapping is a simple bit-select from the physical address, and thus requires minimal additions. This hardware is introduced at every {NoC} injection point.

In summary, the hardware additions proposed in this paper require an additional 4.65 mm\textsuperscript{2} of die area for a system using 16 SPUs. This amounts to a 0.77\% area increase compared to the Marvell ThunderX 2 CPU \mbox{\cite{thunderx2}}, a server-class ARM CPU implemented in \SI{16}{\nano\meter} hosting \SI{32}{MB} of \mbox{{LLC}}.

\section{Discussion}
\label{sec:discussion}

\noindent \textit{\textbf{Why \juan{a stencil accelerator}?}} 
Because of their large contribution to the overall runtime of HPC workloads, improving the performance and energy-efficiency of stencil computations is critical. A wide body of research~\cite{christen2011patus, gysi2015modesto, datta2009optimization, strzodka2010cache, tang2011pochoir, ragan2013halide, meng2011performance, nguyen20103, henretty2011data, jaeger2012automatic, frigo2007memory, olschanowsky2014study, kamil2005impact, stengel2015quantifying,brandvik2010sblock,fuhrer2018near,armejach2018stencil,waidyasooriya2016opencl,de2018designing,sano2013multi,van2019coherently,singh2019narmada,chi2018soda,nero,li2019pims,waidyasooriya2019multi,cattaneo2015accelerate,yantir2020efficient,wester2014deriving} highlights the need for highly \geraldo{efficient} stencil computations. While more general-purpose solutions based on GPUs, FPGAs, and 3D-stacked memory attempt to trade off generality for performance and efficiency, we show that a careful domain-specific hardware/software co-design can improve the performance and energy efficiency even further, at a much \juan{lower} overhead compared to the existing general-purpose solutions.

\noindent \textit{\textbf{Other workloads for \mech.}} Apart from stencils, the \juan{near-LLC} execution model is well-suited for applications with the following properties: (1) their memory access pattern shows temporal locality, (2) they operate on datasets that exceed the capacity of private caches, and (3) have streaming memory access pattern, and as a result do not benefit from deep cache hierarchies. Two examples of workloads that satisfy these properties are high performance computing (HPC) workloads operating on structured grids \cite{schonherr2011multi, allen1999solving}, and dense linear algebra computations \cite{ltaief2012profiling}. While we study our proposal specifically for stencils, which are one of the most widely used kernels in HPC domain, supporting a wider set of applications is possible by redesigning the SPU pipeline to add support for data-dependent divisions that are present in some other HPC workloads. Together with the MAC operations (that \mech~already supports), this extends \mech~to a wider set of use-cases and applications.

\section{Related Work}\label{sec:related_work}
\glsreset{HBM}
To our knowledge, we present the first work that tightly integrates specialized {computation} units into the last-level cache of a CPU to perform stencil computations. In this section, we succinctly compare prior works against \mech.

Due to the high contribution of stencil computation to the overall execution time of HPC applications, a wide body of research has focused on studying and analyzing stencil computations~\cite{li2019pims,gysi2015modesto, ragan2013halide, nguyen20103, stengel2015quantifying, fuhrer2018near, armejach2018stencil, sano2013multi,van2019coherently,singh2019narmada,chi2018soda,nero,waidyasooriya2019multi,cattaneo2015accelerate,yantir2020efficient,wester2014deriving,christen2011patus, datta2009optimization, strzodka2010cache, tang2011pochoir, meng2011performance, henretty2011data, jaeger2012automatic, frigo2007memory, olschanowsky2014study,kamil2005impact,brandvik2010sblock,waidyasooriya2016opencl,de2018designing}. 
Prior works show the applicability of {four} different processing types in accelerating stencil computations: {(1) Near-memory, (2) CPU, (3) GPU, (4) FPGA.}

\noindent \textbf{Near-Memory.} 
\rbc{PIMS~\cite{li2019pims} exploits the high-bandwidth provided by 3D-stacked memories (e.g., \gls{HMC}~\cite{hmc2013hybrid}, \gls{HBM}~\cite{HBM,lee2016simultaneous}) to accelerate stencils. \mech, being a near-LLC accelerator, can be integrated with any commodity processor without {requiring} costly interfacing using through-silicon vias.}

\noindent \textbf{CPU.} Szustak et al.~\cite{szustak2013using} accelerate the MPDATA stencil kernel on a multi-core CPU. Thaler~et al.~\cite{cosmo_knl} ii{use a many-core system to accelerate} weather stencil kernels. 
{Szustak and Bratek\cite{szustak2019performance} propose parametric optimization techniques for the MPDATA application on shared-memory systems.}

\label{lab:R1/10}\Copy{R1/10}{\noindent \textbf{GPU.} GPUs~{\mbox{\cite{anjum2019efficient,fuhrer2018near, sun2021cstuner}}} have been shown to increase performance due to the high degree of parallelism present in the ii{stencil} computation. Wahib et al.~\cite{wahib2014scalable} develop an analytical performance model for choosing an optimal GPU-based execution strategy for various scientific stencil kernels. Gysi et al.~\cite{gysi2015modesto} provide guidelines for optimizing stencil kernels for CPU--GPU systems.}

\label{lab:R1/8}\Copy{R1/8}{\noindent \textbf{FPGA.} More recently, the use of FPGAs to accelerate stencils has been proposed~\cite{van2019coherently,singh2019narmada,chi2018soda,nero,Singh2021trets, sano2013multi, de2021stencilflow, sohrabizadeh2022autodse, wang2017comprehensive, reggiani2021enhancing, koraei2019dcmi, tian2022sasa}. Augmenting general-purpose cores with specialized FPGA accelerators is a promising approach to enhance overall system performance. {However, designing and deploying FPGA-based stencil accelerators have three inherent drawbacks {compared} to integrating \mech's SPUs near LLC and programming them.} 
{First, data needs to be moved to these off-chip external FPGA-based accelerators. The fraction of the total execution time needed for this data movement is not negligible, and it may become larger if the entire stencil data does not fit in the limited FPGA memory (typically smaller than the host main memory).} 
{Second, taking full advantage of FPGAs for accelerating a workload is not a trivial task, as this requires sufficient FPGA programming skills to map the workload and optimize the design for the FPGA microarchitecture.} 
{Third, bitstream generation is a very time-consuming process, especially for high-end FPGAs.} 
{In contrast, \mech~integrates compute units close to the LLC, which avoids unnecessary data movement to an external accelerator. 
\mech~programming is easier and faster than FPGA programming, even if high-level synthesis tools (e.g., OpenCL}~\cite{huang2019analysis, jiang2020boyi}) are used, because \mech~only needs a small number of API functions (\autoref{table:api}{) and does not require time-consuming bitstream generation process.}}

\section{Conclusion}

We present \mech, {the first near-cache acceleration mechanism for stencil computations. \mech~enables high performance and energy-efficient execution of stencil computations by (1)~placing throughput-optimized stencil processing units near the {last-level cache (LLC)} and eliminating  the need to move data to the processor, (2)~orchestrating data accesses to minimize data movement within the cache hierarchy, and (3)~maximizing the utilization of the {LLC} bandwidth. \mech~achieves this with an area overhead of less than 1\% ii{for 16 SPUs in a Marvell ThunderX 2 \mbox{\cite{thunderx2}}, a server-class ARM CPU}. We evaluate \mech~using six widely used stencil kernels with up-to 3-dimensional grid domains. We show that \mech~outperforms a commercial multi-core CPU, on average, by $1.65\times$ ({up to $4.16\times$}) and reduces the energy consumption by $35\%$ ({up to $65\%$}). Compared to a state-of-the-art GPU, \mech~improves performance-per-area\gs{, on average,} by $37\times$ ({up to $190\times$}). We conclude that \mech~is a promising near-cache-processing mechanism for accelerating stencil computations and addressing the memory bottleneck for such computations. We believe and hope that future work builds on \mech~to further ease accelerating important high-performance computing applications that use stencil computations.}

\section{Acknowledgments}

We thank the SAFARI Research Group members for valuable feedback and the stimulating intellectual environment they provide. We acknowledge support from the SAFARI Research Group’s industrial partners, especially ASML, Google, Huawei, Intel, Microsoft, VMware, Xilinx, and the Semiconductor Research Corporation. 
This research was partially supported by the ETH Future Computing Laboratory. 
This research was also partially supported by ACCESS – AI Chip Center for Emerging Smart Systems, sponsored by InnoHK funding, Hong Kong SAR. 

\bibliographystyle{IEEEtran}
\bibliography{refs}

 \phantomsection

\begin{figure}[!h]
\begin{minipage}{\dimexpr 0.5\textwidth-2\fboxsep-2\fboxrule}
\abovecaptionskip=0pt
\sbox0{\includegraphics[width=1in,height=1.25in,clip,keepaspectratio]{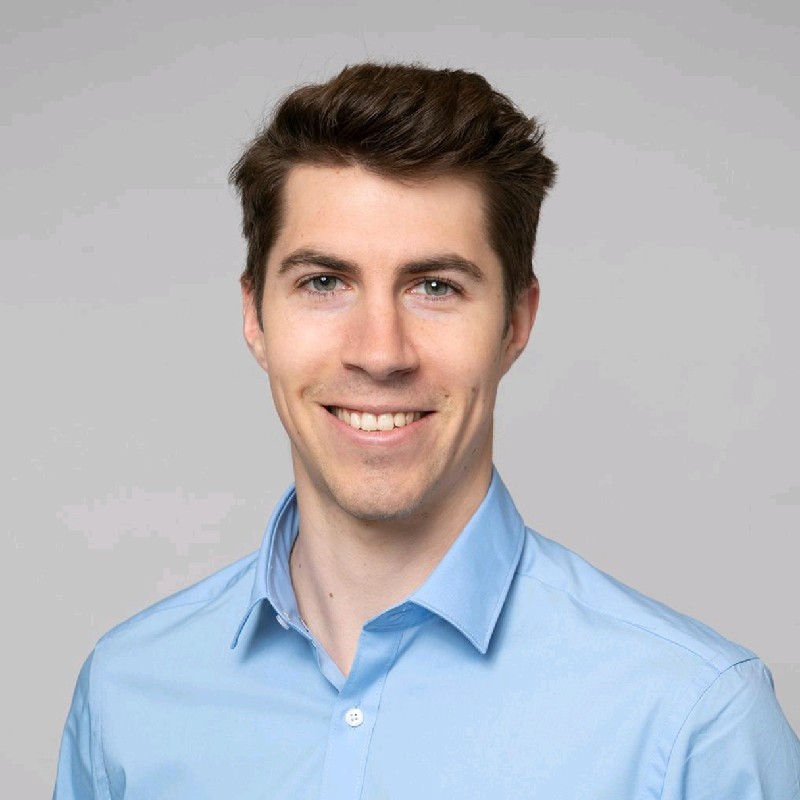}}
\hangafter \numexpr -\ht0/\baselineskip -2\relax
\hangindent \dimexpr \wd0 + 1em\relax
\vbox{\raisebox{-\height}[0pt][0pt]{\box0}}\vspace{-1ex}
\uppercase{Alain Denzler} received a B.S. and an M.S. degree in computer science from ETH Zürich in 2017 and 2020, respectively. He currently works as a software engineer at NVIDIA Switzerland.
\end{minipage}
\end{figure}

\vspace{60pt}
\begin{figure}[!h]
\begin{minipage}{\dimexpr 0.5\textwidth-2\fboxsep-2\fboxrule}
\abovecaptionskip=0pt
\sbox0{\includegraphics[width=1in,height=1.25in,clip,keepaspectratio]{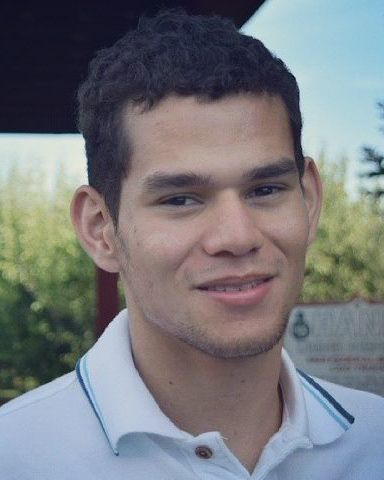}}
\hangafter \numexpr -\ht0/\baselineskip -1\relax
\hangindent \dimexpr \wd0 + 1em\relax
\vbox{\raisebox{-\height}[0pt][0pt]{\box0}}\vspace{-1ex}
\uppercase{Geraldo F. Oliveira} received a B.S. degree in computer science from the Federal University of Viçosa, Viçosa, Brazil, in 2015, and an M.S. degree in computer science from the Federal University of Rio Grande do Sul, Porto Alegre, Brazil, in 2017. Since 2018, he has been working toward a Ph.D. degree with Onur Mutlu at ETH Zürich, Zürich, Switzerland. His current research interests include system support for processing-in-memory and processing-using-memory architectures, data-centric accelerators for emerging applications, approximate computing, and emerging memory systems for consumer devices. He has several publications on these topics.
\end{minipage}
\end{figure}
\vspace{40pt}

\begin{figure}[!h]
\begin{minipage}{\dimexpr 0.5\textwidth-2\fboxsep-2\fboxrule}
\abovecaptionskip=0pt
\sbox0{\includegraphics[width=1in,height=1.25in,clip,keepaspectratio]{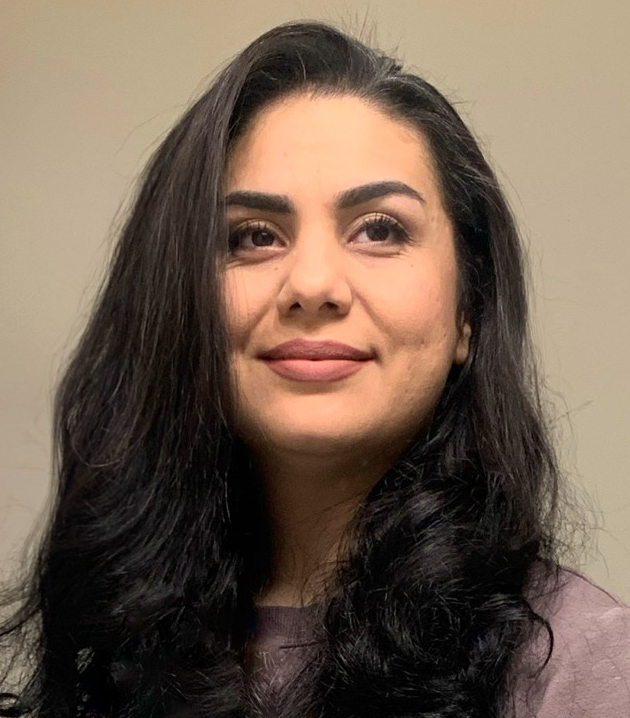}}
\hangafter \numexpr -\ht0/\baselineskip -1\relax
\hangindent \dimexpr \wd0 + 1em\relax
\vbox{\raisebox{-\height}[0pt][0pt]{\box0}}\vspace{-1ex}
\uppercase{Nastaran Hajinazar} is a Senior Researcher at ETH Zürich. Nastaran received her M.S. degree in computer hardware engineering from Sharif University of Technology, Tehran, Iran, in 2011 and her Ph.D. degree in computer science from Simon Fraser University, British Columbia, Canada, in 2020. Her research incorporates several aspects of computer architecture with a significant focus on designing efficient high-performance computing systems, memory architectures, and intelligent memory management techniques.
\end{minipage}
\end{figure}

\begin{figure}[!t]
\begin{minipage}{\dimexpr 0.5\textwidth-2\fboxsep-2\fboxrule}
\abovecaptionskip=0pt
\sbox0{\includegraphics[width=1in,height=1.25in,clip,keepaspectratio]{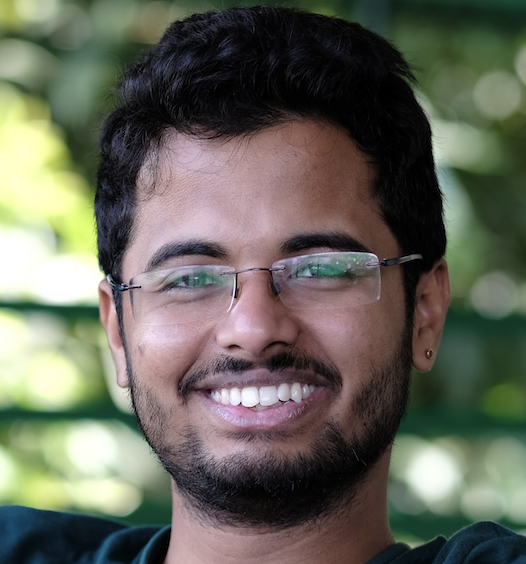}}
\hangafter \numexpr -\ht0/\baselineskip -1\relax
\hangindent \dimexpr \wd0 + 1em\relax
\vbox{\raisebox{-\height}[0pt][0pt]{\box0}}\vspace{-1ex}
\uppercase{Rahul Bera} is a third year Ph.D. student in ETH Zürich Switzerland. His research interests focus on the broad area of memory hierarchy design and applied machine learning in computer architecture. He received his masters degree in computer science from Indian Institute of Technology, Kanpur on 2017. He has previously worked in AMD and Intel Labs, India. 
\end{minipage}
\end{figure}

\begin{figure}[!h]
\begin{minipage}{\dimexpr 0.5\textwidth-2\fboxsep-2\fboxrule}
\abovecaptionskip=0pt
\sbox0{\includegraphics[width=1in,height=1.25in,clip,keepaspectratio]{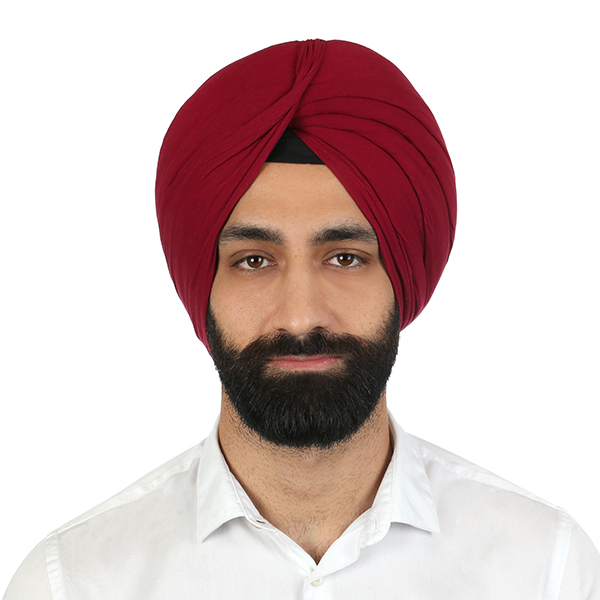}}
\hangafter \numexpr -\ht0/\baselineskip -1\relax
\hangindent \dimexpr \wd0 + 1em\relax
\vbox{\raisebox{-\height}[0pt][0pt]{\box0}}\vspace{-1ex}
\uppercase{Gagandeep Singh} is a Senior Researcher at ETH Zürich. In March 2021, he received his Ph.D. degree from Technische Universiteit Eindhoven, Netherlands, under the supervision of Prof. Henk Corporaal and Prof. Onur Mutlu.  He is passionate about computer architecture, FPGA acceleration, processing-in-memory, bioinformatics,  and machine learning. He obtained a joint M.Sc. degree with distinction in Integrated Circuit Design from Technische Universität München (TUM), Germany, and Nanyang Technological University (NTU), Singapore, in 2017. During his Ph.D.  from June 2018 to January 2020, he was a Predoctoral Researcher at IBM Research Zürich, Switzerland. He has previously worked in Oracle, India, and also performed research at IMEC, Belgium. 
\end{minipage}
\end{figure}

\begin{figure}[!h]
\begin{minipage}{\dimexpr 0.5\textwidth-2\fboxsep-2\fboxrule}
\abovecaptionskip=0pt
\sbox0{\includegraphics[width=1in,height=1.25in,clip,keepaspectratio]{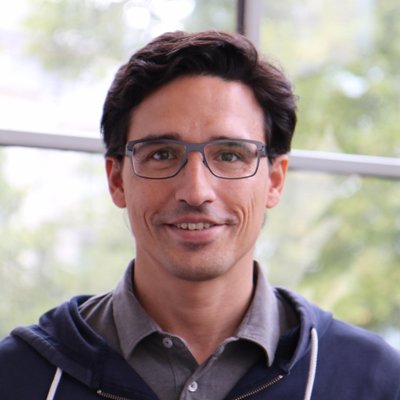}}
\hangafter \numexpr -\ht0/\baselineskip -1\relax
\hangindent \dimexpr \wd0 + 1em\relax
\vbox{\raisebox{-\height}[0pt][0pt]{\box0}}\vspace{-1ex}
\uppercase{Juan Gómez-Luna} is a senior researcher and lecturer at SAFARI Research Group @ ETH Zürich. He received the BS and MS degrees in Telecommunication Engineering from the University of Sevilla, Spain, in 2001, and the PhD degree in Computer Science from the University of Córdoba, Spain, in 2012. Between 2005 and 2017, he was a faculty member of the University of Córdoba. His research interests focus on processing-in-memory, memory systems, heterogeneous computing, and hardware and software acceleration of medical imaging and bioinformatics. He is the lead author of PrIM (https://github.com/CMU-SAFARI/prim-benchmarks), the first publicly-available benchmark suite for a real-world processing-in-memory architecture, and Chai (https://github.com/chai-benchmarks/chai), a benchmark suite for heterogeneous systems with CPU/GPU/FPGA.
\end{minipage}
\end{figure}

\begin{figure}[!t]
\begin{minipage}{\dimexpr 0.5\textwidth-2\fboxsep-2\fboxrule}
\abovecaptionskip=0pt
\sbox0{\includegraphics[width=1in,height=1.25in,clip,keepaspectratio]{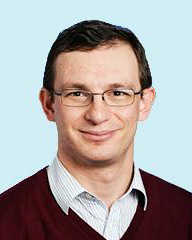}}
\hangafter \numexpr -\ht0/\baselineskip -1\relax
\hangindent \dimexpr \wd0 + 1em\relax
\vbox{\raisebox{-\height}[0pt][0pt]{\box0}}\vspace{-1ex}
\uppercase{Onur Mutlu} is a Professor of Computer Science at ETH Zurich. He is also a faculty member at Carnegie Mellon University, where he previously held the Strecker Early Career Professorship. His current broader research interests are in computer architecture, systems, hardware security, and bioinformatics. A variety of techniques he, along with his group and collaborators, has invented over the years have influenced industry and have been employed in commercial microprocessors and memory/storage systems. He obtained his PhD and MS in ECE from the University of Texas at Austin and BS degrees in Computer Engineering and Psychology from the University of Michigan, Ann Arbor. He started the Computer Architecture Group at Microsoft Research (2006-2009), and held various product and research positions at Intel Corporation, Advanced Micro Devices, VMware, and Google. He received the IEEE High Performance Computer Architecture Test of Time Award, the IEEE Computer Society Edward J. McCluskey Technical Achievement Award, ACM SIGARCH Maurice Wilkes Award, the inaugural IEEE Computer Society Young Computer Architect Award, the inaugural Intel Early Career Faculty Award, US National Science Foundation CAREER Award, Carnegie Mellon University Ladd Research Award, faculty partnership awards from various companies, and a healthy number of best paper or ``Top Pick'' paper recognitions at various computer systems, architecture, and security venues. He is an ACM Fellow, IEEE Fellow for, and an elected member of the Academy of Europe (Academia Europaea). His computer architecture and digital logic design course lectures and materials are freely available on YouTube (https://www.youtube.com/OnurMutluLectures), and his research group makes a wide variety of software and hardware artifacts freely available online (https://safari.ethz.ch/). For more information, please see his webpage at https://people.inf.ethz.ch/omutlu/.
\vspace{180pt}
\end{minipage}
\end{figure}

\newpage
\onecolumn
\section*{APPENDIX}

{This appendix presents some detailed measurements, which correspond to our analyses in Section}~\ref{sec:results}. 
\autoref{tab:app_instr} shows the dynamic instruction count for all evaluated stencils and datasets on the baseline CPU (16 cores) and \mech~(16 SPUs). 
\autoref{tab:app_cycles} shows the number of execution cycles for all evaluated stencils and datasets on the baseline CPU (16 cores), the baseline GPU, and \mech~(16 SPUs). \autoref{tab:app_energy} {shows the energy consumption for all evaluated stencils and datasets on the baseline CPU (16 cores) and \mech~(16 SPUs). 
}

\begin{table*}[h]
\caption{{Dynamic Instruction Count for the Baseline CPU (16 cores) and \mech~(16 SPUs)}}
\vspace{-1mm}
\resizebox{1.0\linewidth}{!}{
\begin{tabular}{l|ccc|ccc|ccc|ccc|ccc|ccc}
\hline
& \multicolumn{3}{c|}{\textbf{Jacobi 1D}} & \multicolumn{3}{c|}{\textbf{7-point 1D}} & \multicolumn{3}{c|}{\textbf{Jacobi 2D}} & \multicolumn{3}{c|}{\textbf{Blur 2D}} & \multicolumn{3}{c|}{\textbf{7-point 3D}} & \multicolumn{3}{c}{\textbf{33-point 3D}} \\
& L2 & LLC & DRAM & L2 & LLC & DRAM & L2 & LLC & DRAM & L2 & LLC & DRAM & L2 & LLC & DRAM & L2 & LLC & DRAM \\
\hline
\hline
\textbf{CPU (16 cores)} & 165840 & 1312867 & 5245651 & 297277 & 2361924 & 9440116 & 537100 & 4311784 & 17255191 & 1804260 & 16552680 & 66329169 & 736767 & 6083864 & 24330380 & 2452622 & 20958248 & 83845023 \\
\textbf{\mech~(16 SPUs)} & 3106 & 23038 & 3034882 & 26470 & 211402 & 3422962 & 5482 & 186718 & 12640918 & 38350 & 337858 & 4135498 & 20002 & 198730 & 21826798 & 261562 & 1050790 & 9321778 \\
\hline
\end{tabular}
}
\label{tab:app_instr}
\vspace{3mm}
\end{table*}

\begin{table*}[h]
\caption{{Execution Cycles for the Baseline CPU (16 cores), the Baseline GPU, and \mech~(16 SPUs)}}
\vspace{-1mm}
\resizebox{1.0\linewidth}{!}{
\begin{tabular}{l|ccc|ccc|ccc|ccc|ccc|ccc}
\hline
& \multicolumn{3}{c|}{\textbf{Jacobi 1D}} & \multicolumn{3}{c|}{\textbf{7-point 1D}} & \multicolumn{3}{c|}{\textbf{Jacobi 2D}} & \multicolumn{3}{c|}{\textbf{Blur 2D}} & \multicolumn{3}{c|}{\textbf{7-point 3D}} & \multicolumn{3}{c}{\textbf{33-point 3D}} \\
& L2 & LLC & DRAM & L2 & LLC & DRAM & L2 & LLC & DRAM & L2 & LLC & DRAM & L2 & LLC & DRAM & L2 & LLC & DRAM \\
\hline
\hline
\textbf{CPU (16 cores)} & 13358 & 95251 & 3838447 & 14702 & 125138 & 5715526 & 26457 & 178032 & 8720011 & 95428 & 742734 & 22729495 & 39029 & 296436 & 7986968 & 115884 & 1009021 & 9060219 \\
\textbf{GPU} & 4030 & 36134 & 135360 & 4108 & 36594 & 139320 & 4646 & 37248 & 140160 & 6950 & 41318 & 153480 & 5184 & 36633 & 140856 & 6758 & 52491 & 278784 \\
\textbf{\mech~(16 SPUs)} & 4569 & 33220 & 4370993 & 8449 & 66393 & 4514872 & 7658 & 58734 & 3931701 & 55764 & 446300 & 5454431 & 29572 & 286675 & 6784185 & 100243 & 1385955 & 13420984 \\
\hline
\end{tabular}
}
\label{tab:app_cycles}
\vspace{3mm}
\end{table*}

\begin{table*}[h]
\caption{{Energy Consumption (J) for the Baseline CPU (16 cores) and \mech~(16 SPUs)}}
\vspace{-1mm}
\resizebox{1.0\linewidth}{!}{
\begin{tabular}{l|ccc|ccc|ccc|ccc|ccc|ccc}
\hline
& \multicolumn{3}{c|}{\textbf{Jacobi 1D}} & \multicolumn{3}{c|}{\textbf{7-point 1D}} & \multicolumn{3}{c|}{\textbf{Jacobi 2D}} & \multicolumn{3}{c|}{\textbf{Blur 2D}} & \multicolumn{3}{c|}{\textbf{7-point 3D}} & \multicolumn{3}{c}{\textbf{33-point 3D}} \\
& L2 & LLC & DRAM & L2 & LLC & DRAM & L2 & LLC & DRAM & L2 & LLC & DRAM & L2 & LLC & DRAM & L2 & LLC & DRAM \\
\hline
\hline
\textbf{CPU (16 cores)} & 0.00012 & 0.00113 & 0.2631221 & 0.000144 & 0.00145 & 0.28253 & 0.000256 & 0.002 & 0.3483945 & 0.0009 & 0.0075 & 0.64639877 & 0.000386 & 0.003364 & 0.469465 & 0.0011542 & 0.010266 & 0.4424779 \\
\textbf{\mech~(16 SPUs)} & 0.000468 & 0.00341 & 0.3114322 & 0.000629 & 0.00469 & 0.59888 & 0.00073 & 0.0055 & 0.8809648 & 0.0015 & 0.0118 & 1.19655244 & 0.001737 & 0.014002 & 1.4752518 & 0.0028739 & 0.027749 & 1.8090142 \\
\hline
\end{tabular}
}
\label{tab:app_energy}
\vspace{3mm}
\end{table*}

\end{document}